%% file: main.tex
\PassOptionsToPackage{table}{xcolor}
\documentclass[11pt]{article}

\usepackage[preprint]{acl}

\usepackage{times}
\usepackage{latexsym}

\usepackage[T1]{fontenc}

\usepackage[utf8]{inputenc}

\usepackage{microtype}

\usepackage{inconsolata}

\usepackage{graphicx} 
\usepackage{luatex85}
\usepackage{threeparttable}
\usepackage{amsmath}
\usepackage{amsfonts}
\usepackage{bm}
\usepackage{subcaption}

\usepackage{inconsolata}
\usepackage{algorithm}
\usepackage{algorithmic}
\usepackage{xcolor}
\usepackage{tcolorbox}
\tcbuselibrary{raster} 
\definecolor{darkblue}{rgb}{0.04, 0.1, 0.35}
\definecolor{lightblue}{rgb}{0.2, 0.6, 0.9}
\definecolor{lightred}{rgb}{0.9, 0.4, 0.4}
\definecolor{lightyellow}{rgb}{0.8, 0.8, 0.35}
\definecolor{lightpurple}{rgb}{0.6, 0.5, 0.7}
\definecolor{lightorange}{rgb}{1.0, 0.5, 0.0}
\definecolor{lightgrey}{rgb}{0.5, 0.5, 0.5}
\usepackage{booktabs} 

\title{\textsc{InvestorBench}: A Benchmark for Financial Decision-Making Tasks with LLM-based Agent}

\author{
\hspace{-1cm}
Haohang Li$^{1,\star}$,
Yupeng Cao$^{1,\star}$,
Yangyang Yu$^{1,\star}$, 
Shashidhar Reddy Javaji$^{1}$,
Zhiyang Deng$^{1}$, \\
\bf
\hspace{-1cm}
Yueru He$^{2}$,
Yuechen Jiang$^{1}$,
Zining Zhu$^{1}$,\,
Koduvayur Subbalakshmi$^{1}$,
Guojun Xiong$^{3}$,\\
\bf
\hspace{-1cm}
Jimin Huang$^{4}$,
Lingfei Qian$^{4}$,
Xueqing Peng $^{4}$,
Qianqian Xie$^{4,\dagger}$,
Jordan W. Suchow$^{1}$\\
\hspace{-1cm}
$^{1}$Stevens Institute of Technology\quad
$^{2}$Columbia University\quad
$^{3}$Harvard University\quad
$^{4}$The Fin AI\\
\hspace{-1cm}
$^{\star}$These authors contributed equally\quad
$^{\dagger}$ Corresponding author: \texttt{qianqian.xie@yale.edu}
}

\begin{document}
\maketitle
\begin{abstract}
Recent advancements have underscored the potential of large language model (LLM)-based agents in financial decision-making. Despite this progress, the field currently encounters two main challenges: (1) the lack of a comprehensive LLM agent framework adaptable to a variety of financial tasks, and (2) the absence of standardized benchmarks and consistent datasets for assessing agent performance. To tackle these issues, we introduce \textsc{InvestorBench}, the first benchmark specifically designed for evaluating LLM-based agents in diverse financial decision-making contexts. \textsc{InvestorBench} enhances the versatility of LLM-enabled agents by providing a comprehensive suite of tasks applicable to different financial products, including single equities like stocks, cryptocurrencies and exchange-traded funds (ETFs). Additionally, we assess the reasoning and decision-making capabilities of our agent framework using thirteen different LLMs as backbone models, across various market environments and tasks. Furthermore, we have curated a diverse collection of open-source, multi-modal datasets and developed a comprehensive suite of environments for financial decision-making.  This establishes a highly accessible platform for evaluating financial agents' performance across various scenarios.
\end{abstract}

\input{intro}
\input{preliminary}

\section{InvestorBench}
\label{sec:def}

He we introduce the detailed architecture of InvestorBench, as illustrated in Figure~\ref{fig:agent-structure}.

\subsection{Benchmark Composition}
\textsc{InvestorBench} is organized into four main components: (1) \textbf{Data Sources and Market Environments}: \textsc{InvestorBench} utilizes a wide range of open-source data and incorporates third-party APIs, such as Yahoo Finance and SEC EDGAR, to create a comprehensive, multi-modal market environment data warehouse. (2) \textbf{LLM Agent}: \textsc{InvestorBench} includes an advanced LLM-based agent equipped with modules for Brain, Perception, Profile, Memory, and Action. This agent is enhanced with external tools (such as tabular data readers and API callers) and data operations (including vector database management, information reinforcement, and retrieval). (3) \textbf{Financial Decision-Making Tasks}: \textsc{InvestorBench} offers three distinct financial decision-making tasks, differentiated by their asset types. (4) \textbf{Evaluation Metrics}: The efficacy of all tasks within \textsc{InvestorBench} is evaluated using a set of standard metrics in the quantitative finance field, providing a thorough evaluation of the decision-making capabilities of the LLM-based agent.

\subsection{Trading Environments}
We release three datasets, each curated from diverse sources, to construct tailored financial market environments for specific tasks. Our objective is to address the current gap in evaluation environments for financial decision-making agent frameworks and to offer a fully open platform for the comprehensive assessment of agents across various tasks. Below, we introduce each environment, categorized by task type, detailing its scope and the data sources it incorporates. \\


\noindent\textbf{Stock market environment} integrates information from multiple sources, including: 1) Daily stock open, high, low, close, and volume (OHLCV) data acquired from Yahoo Finance. 2) Summarized insights from company quarterly and annual reports (Form 10-Q and 10-K) downloaded from the SEC EDGAR database. 3) News articles for seven stocks collected daily between 2020-07-01, and 2021-05-06.  The news data for four of these companies—Microsoft Corporation (MSFT), Johnson \& Johnson (JNJ), UVV Corporation (UVV), and Honeywell International Inc. (HON)-are randomly selected from the pool with the most new records (over five hundred) from the open-access dataset provided by Zhou et al. \cite{zhou-etal-2021-trade}, while the news data for the remaining three companies—Tesla, Inc. (TSLA), Apple Inc. (AAPL), and NIO Inc. (NIO)-are obtained from Refinitiv Real-Time News, which primarily contains high-quality news information from Reuters. 4) The sentiment categories (\textit{'positive', 'negative', 'neutral'}) assigned to each news record are generated by gpt-3.5-turbo-0125.\\

\noindent\textbf{Cryptocurrency market environment} encompasses 1) the daily stock open-high-low-close-volume (OHLCV) acquired from CoinMarketCap; 2) the multisource cryptocurrency news data collected from cryptonews, cryptopotato, and cointelegraph\cite{kagglecryptonews}; 3) news spanning from 2023-02-13 to 2023-11-05 collected by \cite{zhou-etal-2021-trade} in daily frequency. 4) The sentiment categories generated by the same means.\\

\noindent\textbf{ETF market environment} is constructed using News-Informed Financial Trend Yield (NIFTY) dataset \cite{saqur2024nifty}. It contains the processed and curated daily news headlines from 2019-07-29 to 2020-09-21 and generated sentiment categories for each news headline.

In experimental use, we divide the dataset according to the date, with the train set used for the warmup phase to establish the memory database, and the test set used for the test phase to evaluate the model performance.


\subsection{Evaluation metrics}
We employ four widely recognized financial metrics to evaluate and compare the investment performance of various LLMs serving as backbones across different tasks: : Cumulative Return (CR) \cite{hull2007risk}, Sharpe Ratio (SR) \cite{sharpe1994sharpe}, Annualized Volatility (AV) \cite{cochrane1988volatility}, and Maximum Drawdown(MDD) \cite{ang2003downside}. Note that CR and the SR are often considered more essential than AV and MDD in evaluating asset trading performance due to their focus on long-term gains and risk-adjusted returns by their definition. Here, we regard these two metrics as primary metrics when evaluating the experiment outcomes. The detailed explanation is in Appendix~\ref{sec:metric}.

\input{allmodel}

\input{experiment}

\section{Related Work}
\label{sec:related}

\subsection{LLM for Financial Domain} 
The rapid developement of general-domain language models (LMs) has stimulated the exploration of financial LMs, such as pre-trained LMs: \textsc{FinBert} \cite{liu2021finbert,yang2020finbert,araci2019finbert,huang2023finbert}, \textsc{FinBERT-MRC} \cite{zhang2023finbert}, \textsc{FLANG} \cite{shah2022flue}, and several financial LLMs: \textsc{FinGPT}\cite{liu2023fingpt}, \textsc{FinMA} \cite{xie2023pixiu}, \textsc{InvestLM} \cite{yang2023investlm}, {BloombergGPT} \cite{wu2023bloomberggpt}, which leverage extensive training on diverse financial datasets (e.g. stock price data, financial news and analyst reports) and adapt the capabilities of LMs to the unique needs of financial applications. Concurrently, the advancement of LLMs has significantly enhanced the development of language-based agent frameworks in the financial sector, such as \textsc{FinMem} \cite{yu2024finmem}, \textsc{FinAgent} \cite{zhang2024finagent} and \textsc{FinRobot} \cite{yang2024finrobot}, characterized by their adaptability and openness. However, variations in framework design, task scope, and data types present challenges in uniformly evaluating the efficacy of LLM agents in financial scenarios.

\subsection{Financial LLM Benchmarks} 
In the realm of financial LLMs, several benchmarks have been developed: FLUE~\cite{shah2022flue} introduces the first comprehensive benchmark with five financial NLP tasks, including sentiment analysis, headline classification, named entity recognition, structure boundary detection, and question answering. Pixiu~\cite{xie2023pixiu} expands this benchmark to include financial document understanding and classification tasks, incorporating multimodal datasets. FinBen~\cite{xie2024finben} encompasses 36 datasets covering 24 financial tasks. Despite these advancements, there remains a notable gap in benchmarks specifically designed for LLM-based agent applications within the financial sector. 

\section{Conclusion}
\label{sec:con}
\textsc{InvestorBench} offers the community two distinct modes of engagement. The first mode allows participants to integrate their fine-tuned LLMs into the \textsc{InvestorBench}'s agent framework to undertake financial decision-making tasks. This setup enables them to benchmark the performance of their models against those previously experimented with by our work. The second mode permits users to directly incorporate the environment and evaluation metrics of \textsc{InvestorBench} into their own designed agents, facilitating a comparative analysis of their agent design's effectiveness. This dual approach provides a flexible framework for testing and enhancing financial decision-making strategies within the \textsc{InvestorBench} ecosystem. 

Future research efforts will expand the benchmark by incorporating additional information modalities, such as audio (e.g., earnings call recordings) and graphs (e.g., K-lines, trade charts), to explore whether these data types can enhance decision-making quality. The foundational agent framework of \textsc{InvestorBench} is designed to seamlessly accommodate these modalities, ensuring that the extended benchmark remains easy to use and scalable.

\section*{Limitation}
First, \textsc{InvestorBench} is currently focusing on single-asset financial decision-making task, without addressing multi-asset tasks such as portfolio management. Second, copyright restrictions on financial domain data may compromise the quality of the datasets we create, potentially limiting the assessment of model performance.

\section*{Ethical Statement}
The authors take full responsibility for the development of \textsc{InvestorBench}, ensuring that the publicly available part in dataset does not contain personal information, and conform to established ethical guidelines. The data are shared under the MIT license, requiring users to adhere to its terms. \textsc{InvestorBench} is intended for academic and educational purposes only and is not a substitute for professional advice. While efforts have been made to ensure its accuracy, the authors and their institutions disclaim liability for any outcomes arising from its use. Users agree to take responsibility for ethical and lawful use and to indemnify the authors and their affiliates against any claims or damages resulting from reliance on this Material.
\bibliography{main}

\appendix
\input{appendix}

\end{document}

%% file: intro.tex
\section{Introduction}

The recent studies on large language model (LLM)-based agents have demonstrated impressive performance across a range of decision-making tasks in complex and open-ended environments spanning various domains \cite{zhang2024llm, guo2024large, eigner2024determinants, wang2024survey}. However, developing agentic frameworks tailored specifically for financial decision-making remains a significant challenge. This complexity arises from the need for agents to acutely discern and prioritize decisive signals, and then make sequentially high-quality decisions within the volatile and multifaceted financial markets, where information varies in time sensitivity and modality.

\begin{figure}[t]
\centering
\includegraphics[trim=0 0 0 0,clip,height=8cm,width=0.5\textwidth]{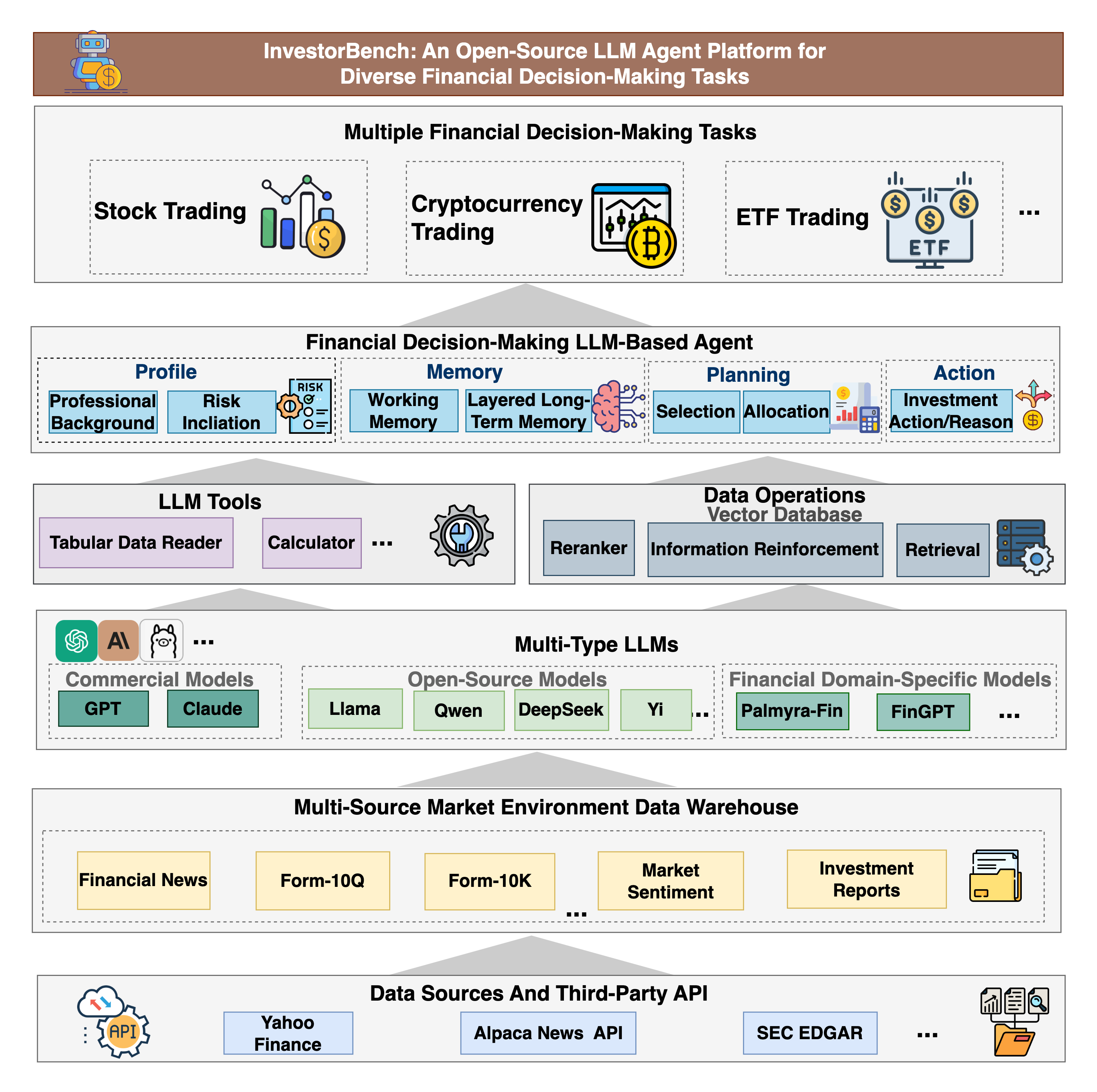}
\caption{General architecture of \textsc{InvestorBench}.}
\label{fig:agent-structure}
\end{figure}

Furthermore, the design of financial agents becomes increasingly complex when applied across multiple decision-making tasks, due to the significant variation in key factors influencing financial decisions across different objectives and task types. For instance, single-equity tasks like stock trading require analyzing company-specific and industry-wide data, including market metrics, sector trends, performance reports, and relevant news \cite{yi2022stock2vec}. In contrast, cryptocurrency trading is highly sensitive to crypto-specific news and sentiment due to its dynamic nature \cite{bhatnagar2023demystifying}. ETFs, on the other hand, typically follow passive investment strategies, emphasizing long-term growth and cost efficiency \cite{madhavan2016exchange}. 

The recent emergence of financial LLM-based agent frameworks such as \textsc{FinMem} \cite{yu2024finmem}, \textsc{FinAgent} \cite{zhang2024finagent}, \textsc{CryptoTrade} \cite{li2024cryptotrade}, \textsc{FinRobot} \cite{yang2024finrobot}, and \textsc{FinCon} \cite{yu2024fincon} has presented a variety of architectural approaches tailored to specific financial tasks. This diversification has sparked substantial interest across both academic and industrial landscapes. \textsc{FinRobot} is engineered specifically for market analysis, while \textsc{FinMem} and \textsc{FinAgent} are oriented towards trading individual equities like stocks and ETFs. \textsc{CryptoTrade} focuses solely on cryptocurrency trading. \textsc{FinCon} pioneers in addressing portfolio management, although it currently handles only compact portfolios consisting of three stock assets. While these frameworks are effective within their respective niches, they generally focus on addressing only limited types of financial decision-making tasks. This restricts them from further demonstrating the broader applicability of these frameworks and limits the comprehensive, comparative insights that could be drawn from their overall decision-making performance. Furthermore, the frequent reliance on proprietary financial data complicates the evaluation of these tools, obscuring their effectiveness and adaptability in broader contexts. Therefore, there is a pressing need to develop innovative benchmarks specifically designed to evaluate LLM-based agents across a wider spectrum of financial decision-making scenarios. Such benchmarks would enable a more robust assessment of these technologies, facilitating advancements that could cater to various financial applications.

We introduce \textsc{InvestorBench}, an open-source, LLM-based agent benchmark that generalizes across a broad range of financial decision-making tasks. Its detailed structure is illustrated in Figure~\ref{fig:agent-structure}. Further developed upon the foundational framework of \textsc{FinMem} \cite{yu2024finmem}, which focuses on single-stock investment decisions, our benchmark extends the scope to encompass an \textbf{ensemble of diverse financial market environments} for various financial tasks. \textsc{InvestorBench}'s cognitive architecture, similar to \textsc{FinMem}, employs a \textbf{layered memory processing} mechanism with distinct decay rates, enabling the agent to store, retrieve, and consolidate insights and reflections more effectively than the pure similarity-based memory retrieval used in \textsc{FinAgent}. This approach ensures that decisions are informed by timely and impactful data, a capability previously shown effective for single-asset trading. These features reflect how human traders draw sequential decisions upon investment signals from multiple sources and varying time sensitivities, allowing the agent to naturally adapt to complex financial tasks. \textsc{InvestorBench} expands its evaluation beyond the original stock trading tasks to encompass three decision tasks significant in the realm of financial investment: \textbf{stock trading}, \textbf{cryptocurrency trading}, and \textbf{ETF investing}.

In summary, we make three key contributions: 1) We establish \textsc{InvestorBench}, an \textbf{innovative and comprehensive financial agentic benchmark} designed to evaluate the reasoning and sequential decision-making capabilities of LLM-based agents in complex, open-ended financial scenarios. This benchmark provides a realistic perspective for assessing the design and performance of such agents. 2) We provide a \textbf{set of open-source, multi-source market environments} that closely mirror real-world conditions. Furthermore, these environments also serve as a standardized platform for evaluating the decision-making performance of other LLM-based financial agents. 3) We present a \textbf{unified, flexible language-agent framework} that allows finance professionals to conveniently customize assess any LLMs serving as the agent’s reasoning core. In this paper, we conduct a holistic evaluation of 13 LLMs — including recent, competitive, and domain-specific fine-tuned models (see Table \ref{tab:allmodel}) to provide a broad overview of their reasoning capabilities in sequential decision-making tasks within financial contexts.



%% file: preliminary.tex
\section{LLM Trading Agents}
In this section, we define a framework of the LLM-based agents in the \textsc{InvestorBench} and formalize the financial decision-making tasks within the context of partially observable Markov decision process (POMDP) \cite{bertsekas1996stochastic,liu2020adaptive,kabbani2022deep}.
\subsection{Definition} The LLM-based agent in \textsc{InvestorBench} is structured as a large language model-modulo framework, designed to match or surpass the capabilities of professional human investors. This framework consists of several interconnected modules, each tailored to handle the distinct challenges presented by the financial market’s volatility and complexity: 


\noindent\textbf{Brain/Backbone (LLM):} This module, which is the LLM itself, serves as the core of the LLM-based agent. It enhances the agent's capabilities by enabling it to understand, process, and generate natural language. This module plays a crucial role in supporting complex decision-making processes, offering interpretations of market-related information, generating predictive analytics, and reflecting on past investment decisions.

\noindent\textbf{Perception:} This module serves a critical function by converting raw market data into a structured format that is compatible with the LLM, specifying what the agent perceives and observes, which includes numerical, textual, and visual information.\\

\noindent\textbf{Profile:} This module serves two functions articulated in natural language. Firstly, it describes the agent's role, highlighting its character as an experienced investor with expert-level knowledge and a self-adaptive risk preference. This risk preference dynamically adjusts based on historical market momentum, allowing the agent to optimize its strategies in real time. Secondly, the module provides a detailed background of the decision-making task, specifying the key characteristics and pertinent information about the target assets involved in the trading decisions, such as equity historical performance, price fluctuations, and sector information. This dual-function module supports the agent's decisions with both the current market context and its historical performance.\\

\begin{figure*}[t]
    \centering
    \includegraphics[width=0.90\textwidth]{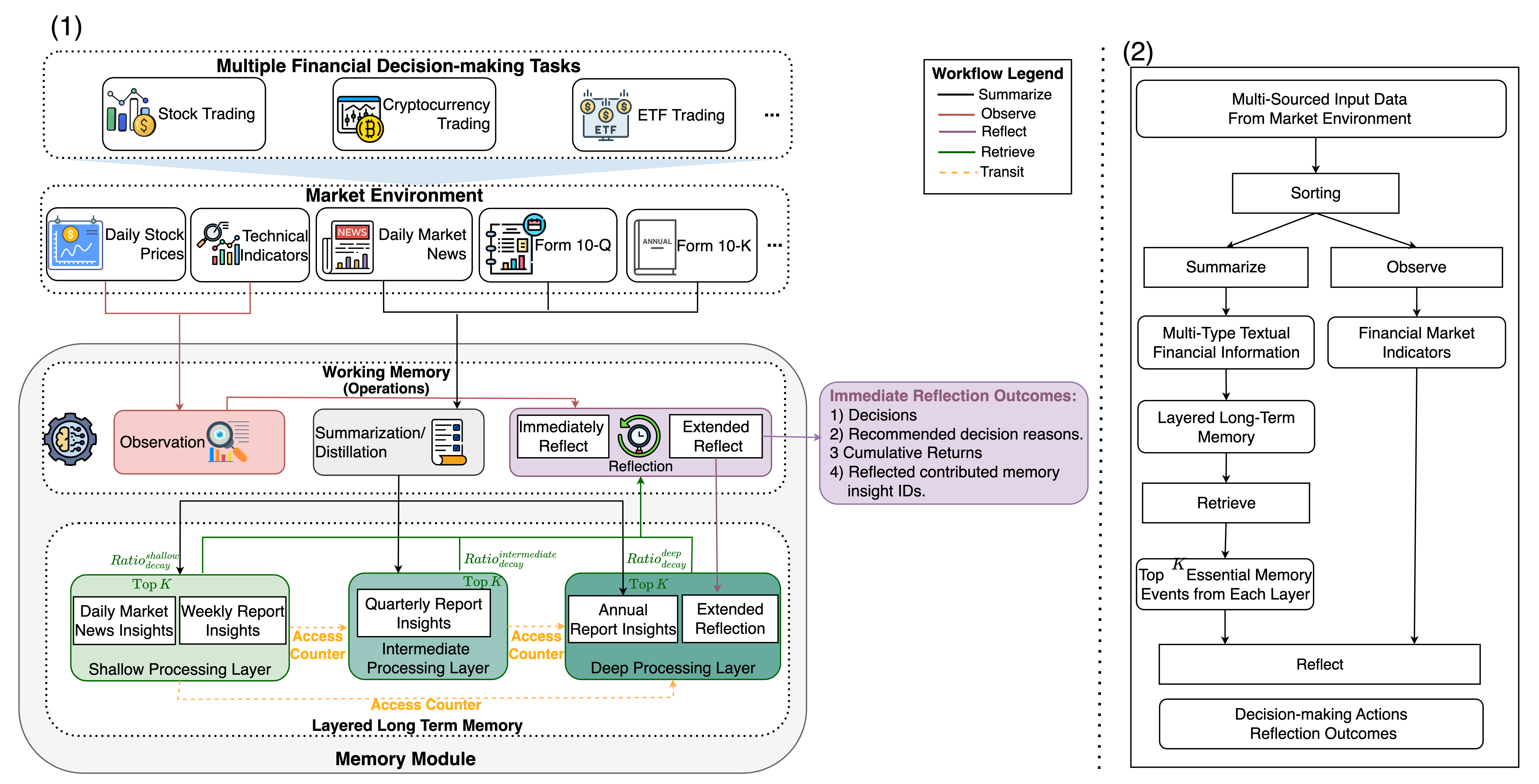}
    \caption{\small (1) The language agent's memory module is crafted to interact with the market environment to conduct various financial decision-making tasks. It contains two core components -- Working Memory and Layered Long-term Memory. (2) The outline of the agent's decision-making workflow for retrieving critical memory events and market observations to inform specific investment decisions.}
    \label{fig:memory_module}
\end{figure*}

\noindent\textbf{Memory:} This module processes and retains essential market data and historical insights, allowing the agent to draw on a rich repository of knowledge for decision-making. Building upon the pioneering work of \citet{yu2024finmem} in \textsc{FinMem}, the memory architecture comprises two primary components: \textbf{Working Memory} and \textbf{Layered Long-Term Memory}, as
depicted in Figure \ref{fig:memory_module}.

\textbf{Working memory:} This component maintains \textsc{FinMem}'s original functionalities: \textit{observation, summarization, and reflection}. It incorporates two reflection mechanisms: \textit{immediate} and \textit{extended}. Immediate reflection produces the agent's reasoning outcomes by integrating current market indicators with the top-$K$ ranked events from each long-term memory layer, which are significant during both warm-up and evaluation stages. In the warm-up stage, the emphasis shifts as the trading direction is predetermined, focusing on understanding market trends and enhancing predictive accuracy. In the evaluation stage, it outputs the trading direction (Buy, Sell, or Hold), the rationale for this decision, identifying the most influential memory events and their respective IDs from each layer.

\textbf{Layered Long-Term Memory:} Inspired by the human cognitive system's varying information decay speeds, Layered Long-Term Memory component structures financial insights across multiple layers. Each layer is represented by a vector database in the Long-Term Memory data warehouse, where information is prioritized and purged based on a specific decay rate. Deeper layers retain information longer with smaller decay rates, while shallower layers, dealing with more transient data, have larger decay rates. This tiered approach is critical as it allows the adaptation of the memory architecture to a broader range of financial tasks beyond single-asset decisions, accommodating an expanded variety of data sources and increasing overall system flexibility. Detailed mechanisms for ranking and decay in each layer are further elaborated in the Appendix~\ref{sec:memory_ranking}.

\noindent\textbf{Action:} This module executes trading and investment decisions based on the analysis provided by other modules. It directly outputs $\{\textit{``Buy",~``Sell",~``Hold"}\}$ for traded asset (stock, crypto, or ETF), as instructed by the backbone LLM. Action module synthesizes the operational outcomes from the Profile and Memory modules to facilitate precise and well-informed investment decisions. For its daily trading operations, the agent can choose from three specific actions for the traded asset: ``Buy", ``Sell", or ``Hold". The functionality and input requirements of this module differ significantly between the warm-up and evaluation stages: \textit{during the warm-up stage}, the agent observes daily adjusted price differences between consecutive days, which are critical for identifying potential ``Buy" or ``Sell" signals. This period allows the agent to calibrate and adjust its decision-making strategies based on near-term market movements; \textit{during the evaluation stage}, access to future price data is restricted, compelling the agent to rely solely on available historical data and its cognitive processing capabilities. In response to trading inquiries, the module integrates historical Profit \& Loss (PnL), outcomes from extended reflections, and the top-$K$ retrieved memories. This integration ensures that each trading decision is grounded in a comprehensive analysis of past performance and current market conditions.

\subsection{Modeling financial decision-making}
Formally, we model a financial decision-making process as infinite horizon POMDP with time index $\mathbb{T}=\{0,1,2,\cdots\}$ and discount factor $\alpha\in(0,1]$. This POMDP contains: (1) a state space $\mathcal{X}\times\mathcal{Y}$ where $\mathcal{X}$ is the observable component and $\mathcal{Y}$ is unobservable component of the financial market; (2) the action space of the agent is $\mathcal{A}$, which is modeled as $\{\textit{``Buy",~``Sell",~``Hold"}\}$; (3) the reward function $\mathcal{R}(o,b,a):\mathcal{X}\times\mathcal{Y}\times\mathcal{A}\to\mathbb{R}$ uses daily profit \& loss (PnL) as the output; (4) the observation process $\{O_t\}_{t\in\mathbb{T}}\subseteq\mathcal{X}$ is a multi-dimensional process (5) the reflection process $\{B_t\}_{t\in\mathbb{T}}\subseteq\mathcal{Y}$ represents the agent's self-reflection, which is updated from $B_t$ to $B_{t+1}$ on daily basis \cite{griffiths2023bayes}; (6) the action $A_t\sim\pi(\cdot|\text{prompt})$ represents the way to make investment decision driven by the language conditioned policy $\pi$. By denoting daily profit \& loss (PnLs) by $R^{\pi}_t=\mathcal{R}(O_t, B_t, A_t)$ and the set of all admissible language conditioned policies as $\Pi=\{\pi(\cdot|\text{prompt})\}$, the optimization objective of financial decision-making task is then:
\begin{equation} \label{objective}
\max_{\pi\in\Pi} \mathbb{E}\Big[\sum_{t\in\mathbb{T}}\alpha^tR^{\pi}_t\Big]
\end{equation}

%% file: allmodel.tex
\begin{table}[t]
    \centering
    \footnotesize
    \caption{\textsc{InvestorBench} evaluates 13 proprietary or open-source LLMs on financial decision-making tasks.}
    \label{tab:allmodel}
    \vspace{-2mm}
    \renewcommand{\tabcolsep}{0.7pt}
    \renewcommand{\arraystretch}{1.1}
    \resizebox{\columnwidth}{!}{
        \begin{tabular}{@{}lccclccc@{}}
            \toprule
            \rowcolor[HTML]{ECF4FF} 
            Model & \#Size & Form & Ver. & Model & \#Size & Form & Ver. \\ 
            \midrule
            \texttt{gpt-4}\protect\cite{achiam2023gpt} & N/A & api & 0613 & 
            \cellcolor[HTML]{ECF4FF}\texttt{Qwen2.5-7b}\protect\cite{Qwen2.5} & 
            \cellcolor[HTML]{ECF4FF}7B & 
            \cellcolor[HTML]{ECF4FF}open & 
            \cellcolor[HTML]{ECF4FF}Instruct \\
            
            \texttt{gpt-4o}\protect\cite{ChatGPT} & N/A & api & 0806 & 
            \cellcolor[HTML]{ECF4FF}\texttt{Qwen2.5-32b}\protect\cite{Qwen2.5} & 
            \cellcolor[HTML]{ECF4FF}32B & 
            \cellcolor[HTML]{ECF4FF}open & 
            \cellcolor[HTML]{ECF4FF}Instruct \\
            
            \texttt{gpt-o1-preview} & N/A & api & 0912 & 
            \cellcolor[HTML]{ECF4FF}\texttt{Qwen2.5-72b}\protect\cite{Qwen2.5} & 
            \cellcolor[HTML]{ECF4FF}72B & 
            \cellcolor[HTML]{ECF4FF}open & 
            \cellcolor[HTML]{ECF4FF}Instruct \\
            
            \cellcolor[HTML]{ECF4FF}\texttt{DeepSeek-v2}\protect\cite{xin2024deepseek} & 
            \cellcolor[HTML]{ECF4FF}15B & 
            \cellcolor[HTML]{ECF4FF}open & 
            \cellcolor[HTML]{ECF4FF}Lite & 
            \texttt{llama3.1-8b}\protect\cite{llama3.1} & 
            8B & 
            open & 
            Instruct \\
            
            \cellcolor[HTML]{ECF4FF}\texttt{DeepSeek-llm}\protect\cite{xin2024deepseek} & 
            \cellcolor[HTML]{ECF4FF}67B & 
            \cellcolor[HTML]{ECF4FF}open & 
            \cellcolor[HTML]{ECF4FF}Chat & 
            \texttt{llama3.1-70b}\protect\cite{llama3.1} & 
            70B & 
            open & 
            Instruct \\
            
            \texttt{Yi-1.5-9b}\protect\cite{young2024yi} & 
            9B & 
            open & 
            Chat & 
            \cellcolor[HTML]{ECF4FF}\texttt{Palmyra-Fin}\protect\cite{Palmyra-Fin-70B-32k} & 
            \cellcolor[HTML]{ECF4FF}70B & 
            \cellcolor[HTML]{ECF4FF}open & 
            \cellcolor[HTML]{ECF4FF}32K \\
            
            \texttt{Yi-1.5-34b}\protect\cite{young2024yi} & 
            34B & 
            open & 
            Chat & 
            \cellcolor[HTML]{ECF4FF}\texttt{}\protect & 
            \cellcolor[HTML]{ECF4FF} & 
            \cellcolor[HTML]{ECF4FF}& 
            \cellcolor[HTML]{ECF4FF} \\
            \bottomrule
        \end{tabular}
    }
    \vspace{-7mm}
\end{table}

%% file: experiment.tex
\section{Experiment and Discussion}
\label{sec:exp}

\begin{table*}[thbp]
\renewcommand{\arraystretch}{1}
\vspace{-0.2cm}
\setlength{\abovecaptionskip}{0.1cm}
\centering
\caption{Performance of stock trading with different LLMs as backbone model across seven stocks.}
\label{tab:stock-trading-performance}
\begin{threeparttable}
\scalebox{0.50}{
\begin{tabular}{@{}lcccc|cccc|cccc|cccc@{}}
\toprule
\textbf{Model} & \multicolumn{4}{c}{\textbf{MSFT}} & \multicolumn{4}{c}{\textbf{JNJ}} & \multicolumn{4}{c}{\textbf{UVV}} & \multicolumn{4}{c}{\textbf{HON}} \\
\cmidrule(lr){2-5}\cmidrule(lr){6-9}\cmidrule(lr){10-13}\cmidrule(lr){14-17}
& \textbf{CR\(\uparrow\)} & \textbf{SR\(\uparrow\)} & \textbf{AV\(\downarrow\)} & \textbf{MDD\(\downarrow\)}
& \textbf{CR\(\uparrow\)} & \textbf{SR\(\uparrow\)} & \textbf{AV\(\downarrow\)} & \textbf{MDD\(\downarrow\)}
& \textbf{CR\(\uparrow\)} & \textbf{SR\(\uparrow\)} & \textbf{AV\(\downarrow\)} & \textbf{MDD\(\downarrow\)}
& \textbf{CR\(\uparrow\)} & \textbf{SR\(\uparrow\)} & \textbf{AV\(\downarrow\)} & \textbf{MDD\(\downarrow\)} \\
\midrule
\textbf{Buy \& Hold}
& 15.340 & 1.039 & 24.980 & 9.428 
& 13.895 & 1.343 & 17.500 & 9.847 
& 36.583 & 2.112 & 29.299 & 15.406 
& 33.256 & 2.347 & 23.967 & 9.195 \\
\midrule
\multicolumn{17}{c}{\textit{\textbf{Financial Domain Models}}} \\
\textbf{Palmyra-Fin-70B}  
& 14.697 & 0.897 & 27.518 & 9.428 
& 5.748 & 0.450 & 19.317 & 9.367 
& 37.875 & 2.039 & 31.200 & 15.967 
& 20.016 & 1.464 & 22.974 & 6.824 \\
\midrule
\multicolumn{17}{c}{\textit{\textbf{Proprietary Models}}} \\
\textbf{GPT-o1-preview} 
& 17.184 & 0.962 & 30.000 & 9.428 
& 13.561 & 1.086 & 20.864 & 9.847 
& 41.508 & 2.147 & 32.479 & 9.633 
& 13.162 & 0.776 & 28.511 & 11.558 \\
\textbf{GPT-4} 
& 16.654 & 0.932 & 30.022 & 9.428 
& 13.712 & 1.103 & 20.894 & 9.860 
& 31.791 & 1.640 & 32.567 & 10.434 
& 34.342 & 2.005 & 28.779 & 9.195 \\
\textbf{GPT-4o} 
& 12.461 & 0.924 & 22.653 & 6.647
& 9.099 & 0.875 & 17.471 & 7.169 
& 8.043 & 0.496 & 27.241 & 14.889 
& 38.540 & 2.418 & 26.782 & 8.979 \\
\midrule
\multicolumn{17}{c}{\textit{\textbf{Open-Source Models}}} \\
\textbf{Qwen2.5-72B-Instruct}  
& 7.421 & 0.588 & 21.238 & 6.973 
& 14.353 & 1.140 & 20.995 & 9.812 
& 37.178 & 1.822 & 34.223 & 13.365 
& 34.309 & 2.000 & 28.779 & 9.292 \\
\textbf{Llama-3.1-70B-Instruct}  
& 17.396 & 1.335 & 21.892 & 7.045 
& 13.868 & 1.121 & 20.779 & 9.825 
& 35.981 & 1.728 & 34.986 & 15.406 
& 43.944 & 2.646 & 27.903 & 8.993 \\
\textbf{DeepSeek-67B-Chat} 
& 13.941 & 0.834 & 28.081 & 7.850 
& 14.426 & 1.185 & 20.450 & 9.825 
& 29.940 & 1.481 & 33.964 & 15.407 
& 32.536 & 1.909 & 28.628 & 10.782 \\
\textbf{Yi-1.5-34B-Chat}  
& 22.093 & 1.253 & 29.613 & 9.428 
& 14.004 & 1.180 & 19.938 & 9.847 
& 20.889 & 1.020 & 34.417 & 14.936 
& 30.743 & 1.823 & 28.335 & 9.195 \\
\textbf{Qwen2.5-32B-Instruct}  
& -0.557 & -0.041 & 22.893 & 8.946 
& 2.905 & 0.292 & 16.725 & 7.169 
& -1.623 & -0.097 & 27.973 & 17.986 
& 26.332 & 1.980 & 22.348 & 5.261 \\
\textbf{DeepSeek-V2-Lite (15.7B)}  
& 11.904 & 0.694 & 28.796 & 16.094 
& -7.482 & -0.670 & 18.773 & 17.806 
& 33.560 & 1.703 & 33.099 & 12.984 
& 16.686 & 0.974 & 28.771 & 16.806 \\
\textbf{Yi-1.5-9B-Chat}  
& 19.333 & 1.094 & 29.690 & 9.428 
& 18.606 & 1.611 & 19.409 & 10.986 
& 49.415 & 2.410 & 34.446 & 11.430 
& 29.028 & 1.700 & 28.682 & 12.588 \\
\textbf{Llama-3.1-8B-Instruct}  
& 22.703 & 1.322 & 28.855 & 7.385 
& 13.988 & 1.486 & 20.460 & 9.969 
& 41.108 & 1.981 & 34.866 & 16.429 
& 39.079 & 2.320 & 28.299 & 10.341 \\
\textbf{Qwen-2.5-Instruct-7B}  
& -10.305 & -0.724 & 23.937 & 23.371 
& 21.852 & 0.980 & 37.425 & 9.573 
& 11.752 & 0.853 & 22.988 & 15.451 
& 4.291 & 0.285 & 24.933 &14.156 \\
\bottomrule
\end{tabular}
}
\end{threeparttable}
\vspace{-0.2cm}
\end{table*}

\begin{table*}[thbp]
\renewcommand{\arraystretch}{1}
\vspace{-0.2cm}
\setlength{\abovecaptionskip}{0.1cm}
\centering
\begin{threeparttable}
\scalebox{0.50}{
\begin{tabular}{@{}lcccc|cccc|cccc|cccc@{}}
\toprule
\textbf{Model} & \multicolumn{4}{c}{\textbf{TSLA}} & \multicolumn{4}{c}{\textbf{AAPL}} & \multicolumn{4}{c}{\textbf{NIO}} & \multicolumn{4}{c}{\textbf{Average}} \\
\cmidrule(lr){2-5}\cmidrule(lr){6-9}\cmidrule(lr){10-13}\cmidrule(lr){14-17}
& \textbf{CR\(\uparrow\)} & \textbf{SR\(\uparrow\)} & \textbf{AV\(\downarrow\)} & \textbf{MDD\(\downarrow\)}
& \textbf{CR\(\uparrow\)} & \textbf{SR\(\uparrow\)} & \textbf{AV\(\downarrow\)} & \textbf{MDD\(\downarrow\)}
& \textbf{CR\(\uparrow\)} & \textbf{SR\(\uparrow\)} & \textbf{AV\(\downarrow\)} & \textbf{MDD\(\downarrow\)}
& \textbf{CR\(\uparrow\)} & \textbf{SR\(\uparrow\)} & \textbf{AV\(\downarrow\)} & \textbf{MDD\(\downarrow\)} \\
\midrule
\textbf{Buy \& Hold}
& 39.244 & 0.869 & 75.854 & 37.975
& 10.837 & 0.470 & 38.984 & 19.119
& 52.216 & 0.858 & 107.197 & 47.766
& 34.099 & 0.732 & 74.012 & 34.953 \\
\midrule
\multicolumn{17}{c}{\textit{\textbf{Financial Domain Models}}} \\
\textbf{Palmyra-Fin-70B}  
& -6.661 & -0.222 & 50.379 & 25.820
& 8.562 & 0.372 & 38.891 & 25.466
& -3.261 & -0.057 & 101.711 & 58.406
& -0.453 & 0.031 & 63.660 & 36.564 \\
\midrule
\multicolumn{17}{c}{\textit{\textbf{Proprietary Models}}} \\
\textbf{GPT-o1-preview} 
& 34.499 & 0.796 & 72.822 & 35.490
& 8.238 & 0.422 & 33.045 & 14.412
& 32.433 & 0.558 & 102.470 & 54.016
& 25.057 & 0.592 & 69.446 & 34.639 \\
\textbf{GPT-4} 
& 45.246 & 1.190 & 63.896 & 25.031
& 9.889 & 0.440 & 38.008 & 19.119
& 75.952 & 1.286 & 104.059 & 37.867
& 43.696 & 0.972 & 68.654 & 27.339 \\
\textbf{GPT-4o} 
& 45.946 & 1.348 & 57.281 & 21.631
& 7.405 & 0.457 & 27.434 & 12.824
& 63.743 & 1.318 & 85.210 & 29.220
& 39.031 & 1.041 & 56.642 & 21.225 \\
\midrule
\multicolumn{17}{c}{\textit{\textbf{Open-Source Models}}} \\
\textbf{Qwen2.5-72B-Instruct}  
& 39.112 & 1.075 & 61.136 & 26.985
& 11.935 & 0.572 & 35.293 & 19.119
& 87.412 & 2.181 & 70.628 & 12.464
& 46.153 & 1.276 & 55.686 & 19.523 \\
\textbf{Llama-3.1-70B-Instruct}  
& 37.545 & 0.891 & 70.815 & 29.813
& 12.772 & 0.583 & 37.057 & 16.021
& 66.522 & 1.118 & 104.848 & 46.379
& 38.946 & 0.864 & 70.907 & 30.738 \\
\textbf{DeepSeek-67B-Chat} 
& 35.647 & 0.885 & 67.660 & 33.359
& 14.213 & 0.666 & 36.118 & 10.876
& 30.963 & 0.599 & 91.146 & 45.855
& 26.941 & 0.717 & 64.975 & 30.030\\
\textbf{Yi-1.5-34B-Chat}  
& 35.364 & 0.808 & 73.561 & 35.490
& 14.227 & 0.623 & 38.596 & 19.432
& 64.307 & 1.063 & 106.597 & 48.042
& 37.966 & 0.831 & 72.918 & 34.321 \\
\textbf{Qwen2.5-32B-Instruct}  
& 21.336 & 0.729 & 49.157 & 20.704
& 13.220 & 1.089 & 20.522 & 8.943
& 28.096 & 0.652 & 72.344 & 37.975
& 20.884 & 0.823 & 47.341 & 22.541 \\
\textbf{DeepSeek-V2-Lite (15.7B)}  
& 31.458 & 0.744 & 68.524 & 35.404
& 27.016 & 1.221 & 11.860 & 37.435
& 27.762 & 0.474 & 103.193 & 48.478
& 28.745 & 0.813 & 61.192 & 40.439 \\
\textbf{Yi-1.5-9B-Chat}  
& 31.350 & 0.703 & 74.895 & 37.975
& 3.640 & 0.162 & 37.947 & 17.578
& 33.748 & 0.569 & 104.502 & 55.284
& 22.913 & 0.478 & 72.448 & 36.946 \\
\textbf{Llama-3.1-8B-Instruct}  
& 35.622 & 0.832 & 71.936 & 36.383
& 7.079 & 0.309 & 38.757 & 18.747
& 33.689 & 0.560 & 106.094 & 56.527
& 25.463 & 0.567 & 72.262 & 37.219 \\
\textbf{Qwen-2.5-Instruct-7B}  
& 41.203 & 0.925 & 74.862 & 37.975
& 14.336 & 0.694 & 34.924 & 12.029
& 33.007 & 0.546 & 106.437 & 53.054
& 29.515 & 0.722 & 72.074 & 34.353 \\
\bottomrule
\end{tabular}
}
\begin{tablenotes}
\footnotesize
\item[1] \small{The Buy \& Hold strategy is a passive investment approach commonly used as a baseline strategy, where an investor\\
purchases stocks and holds onto them for an extended period regardless of market fluctuations.}
\item[2] \small{An upward arrow (\(\uparrow\)) next to a metric indicates that higher values signify better performance, while a downward arrow (\(\downarrow\))} \\
indicates that lower values are preferable.
\item[3] \small{The numbers highlighted in red indicate the best-performing outcomes for the corresponding metrics.}
\end{tablenotes}
\end{threeparttable}
\vspace{-0.2cm}
\end{table*}

\begin{figure}[h]
    \centering
    \begin{subfigure}[b]{0.45\textwidth}
        \centering
        \includegraphics[width=\textwidth]{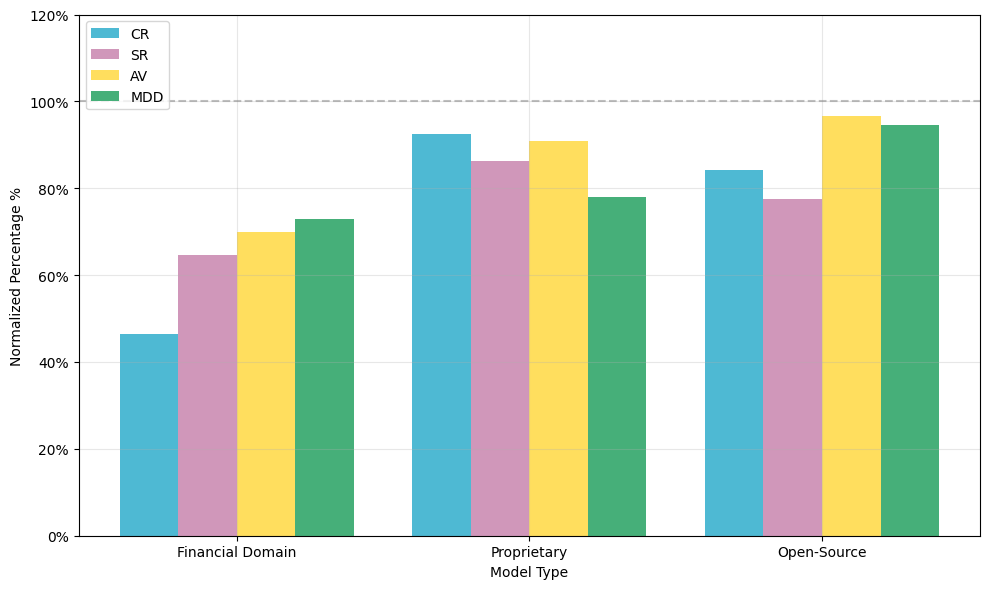}
        \caption{The performance comparison by different model types.}
        \label{fig:comparison-type}
    \end{subfigure}
    \hfill 
    \begin{subfigure}[b]{0.45\textwidth}
        \centering
        \includegraphics[width=\textwidth]{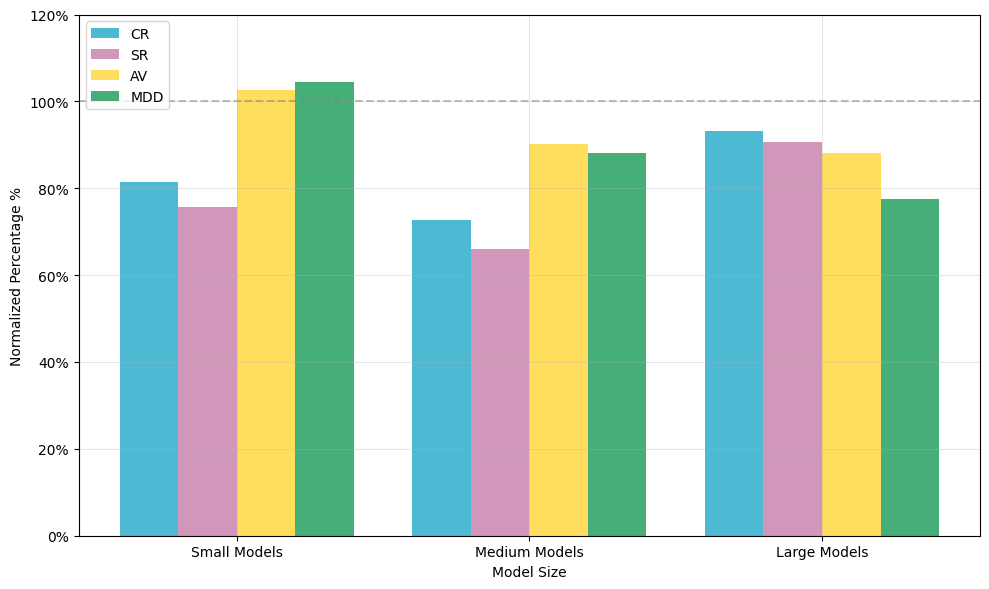}
        \caption{By model parameter sizes across open-source LLMs. \small{\textit{Note: Small-size models refer to models with no more than 10B parameters. Medium-size models refer to models with parameter numbers in the range of (10B, 65B]. Large-size models are those with more than 65B parameters.}}}
        \label{fig:comparison-size}
    \end{subfigure}
    \caption{Agent Performance Comparisons from two key perspectives. \small{\textit{The CR, SR, AV, and MDD represent the average values for each model type, expressed as a percentage relative to the Buy \& Hold strategy.}}}
\end{figure}
To establish a baseline and assess the performance of LLM agents, we standardize experimental settings and evaluation metrics across various financial decision-making tasks. Results are presented on a task-by-task basis. We report the performance of \textsc{InvestorBench} on three single-asset trading tasks: \textbf{\textit{stocks}}, \textbf{\textit{cryptocurrencies}}, and \textbf{\textit{ETFs}} trading, using closed-source, open-source, and domain-specific LLMs.

\subsection{Experiment Setup}
Table~\ref{tab:allmodel} summarizes the performance of a comprehensive list of trading agents.
For single equity tasks, the baseline is set up by Buy and Hold strategy, while for portfolio management task, it is set up by an equal-weight portfolio with the detailed rational explained in Appendix. 
In our experiments, the temperature parameter of all LLM-based agent systems is set at 0.6 to balance response consistency and reasoning creativity.  The performance metrics are reported for the test trajectory with the median CR, SR, AV, and MDD from five repeated epochs. (If the median of these metrics does not belong to the same epoch, the performance is based on the trajectory with the median SR.) 

Furthermore, the selection of warm-up and test periods differs across various tasks due to the varying time spans of data collected to construct the agent environment. For the single-asset trading tasks, the warm-up period of stock trading is from 2020-07-01 to 2020-09-30 and the test period is from 2020-10-01 to 2021-05-06. The warm-up period of cryptocurrency trading is from 2023-02-11 to 2023-04-04 and the test period is from 2023-04-05 to 2023-11-05. The warm-up period of ETF trading is from 2019-07-29 to 2019-12-30 and the test period is from 2020-01-02 to 2020-09-21. 

For LLM deployment, we utilize vllm to deploy LLMs. For small-scale LLMs (under 10B parameters), we deploy models on two RTX A6000 GPUs, each with 48GB DRAM. For mid-scale LLMs (10B to 65B parameters), we use four RTX A6000 GPUs. For large-scale LLMs (over 65B parameters), models are deployed on eight A100 GPUs, each equipped with 80GB DRAM.



\subsection{Result 1: Stock Trading}

Table~\ref{tab:stock-trading-performance} presents the performance of thirteen backbone models across seven stocks, accompanied by the average of each metric for all stocks to offer a more comprehensive view of their overall performance. We outline three key insights as follows:

\noindent\textbf{Superior stock trading performance is achieved with proprietary LLMs as agent backbones} Compared to agents employing open-source or financial-domain-specific fine-tuned LLMs, those using the three proprietary LLMs demonstrated significantly higher and more consistent average CR and SR, as shown in Figure~\ref{fig:comparison-type}. Despite being fine-tuned with extensive financial contexts, domain-specific LLMs did not provide a decisive advantage in sequential stock trading decision-making tasks. This may be attributed to their primary training for other functions, such as long financial report analysis exemplified by Palmyra-Fin-70B, rather than decision-making. 

\noindent\textbf{Model parameter size increment enhances agent financial decision-making quality and robustness.} In the category of open-source LLMs, those exceeding 67B parameters displayed superior CRs and SRs, along with markedly less variance within their category, as illustrated in Figure~\ref{fig:comparison-size} and Table~\ref{tab:stock-trading-performance}. This underscores the prevailing belief that the reasoning capabilities of LLMs are proportionate to their parameter size, which holds also true in stock trading, which is a sequential decision-making task in an open-ended, volatile environment by nature.

\noindent\textbf{Proprietary models exhibit significantly stronger decision-making capabilities compared to even the largest open-source LLMs under complex, mixed market conditions,} though this advantage is less evident in relatively monotone market environments. During the test phase, primarily influenced by the range of open-source data collected, TSLA and NIO exhibited volatility with mixed upward and downward stock price trends, whereas the other five stocks generally showed bullish trends. The investment signals derived from such complex markets tend to be noisy or delayed, as illustrated in Appendix~\ref{sec:example_tsla}. We observed that proprietary models possess a superior ability to manage these challenging conditions and consistently deliver better performance outcomes than large-sized open-source LLMs. Their reasoning capability enables them to effectively utilize other decision-relevant information, such as historical momentum, current holdings, and, most critically, self-reflection outcomes from the agents, thereby facilitating more accurate decisions.  

\subsection{Result 2 \& 3: Cryptocurrency Trading and ETF Trading}

In the test phases of both cryptocurrency and ETF trading tasks, market trends are mixed. Notably, the cryptocurrency task shows significantly smaller price fluctuations compared to the ETF task. We outline the key features of using an LLM-agent to make financial decisions across these two distinct markets as follows:

\noindent\textbf{Large-sized open-source models and proprietary models are needed to effectively capture trading signals of cryptocurrency markets, which are highly sensitive to news and financial sentiment.} As shown in Table~\ref{tab:crypto-trading-performance}, using mid-sized and small-sized open-source models as the decision-making agent backbone generally results in weaker performance than the market baseline with respect to CR and SR.

\noindent\textbf{ETF investment requires proprietary models enriched with extensive pre-trained knowledge to serve as the agent’s “brain” and provide robust reasoning support.} As shown in Table~\ref{etf_result}, proprietary models significantly outperform open-source and financial domain-specific models in this task. This advantage arises from the complexity of ETF trading, which necessitates interpreting actionable signals across diverse sectors, demanding more strategic, long-term decisions grounded in deep comprehension and reflection anchored by rich pre-contexts.

\begin{table}[t]
\centering
\caption{Performance of cryptocurrency trading with different LLMs as backbone models across Bitcoin (BTC) and Ethereum (ETH).}
\label{tab:crypto-trading-performance}
\begin{threeparttable}
\scalebox{0.48}{  
\begin{tabular}{@{}lcccc|cccc@{}}  
\toprule
\textbf{Model} & \multicolumn{4}{c}{\textbf{BTC}} & \multicolumn{4}{c}{\textbf{ETH}} \\
\cmidrule(lr){2-5} \cmidrule(lr){6-9}
& \textbf{CR\(\uparrow\)} & \textbf{SR\(\uparrow\)} & \textbf{AV\(\downarrow\)} & \textbf{MDD\(\downarrow\)} & \textbf{CR\(\uparrow\)} & \textbf{SR\(\uparrow\)} & \textbf{AV\(\downarrow\)} & \textbf{MDD\(\downarrow\)} \\
\midrule
\textbf{Buy \& Hold} & 21.821 & 0.683 & 37.426 & 20.796 & 4.528 & 0.146 & 41.817 & 29.889 \\
\midrule
\multicolumn{9}{c}{\textit{\textbf{Financial Domain Models}}} \\
\textbf{Palmyra-Fin-70B} & -20.812 & -1.212 & 20.036 & 27.782 & 4.795 & 0.240 & 26.924 & 16.405 \\
\midrule
\multicolumn{9}{c}{\textit{\textbf{Proprietary Models}}} \\
\textbf{GPT-o1-preview} & 34.060 & 1.114 & 35.846 & 17.075 & 2.496 & 0.085 & 39.641 & 27.692 \\
\textbf{GPT-4} & 22.396 & 0.828 & 31.699 & 17.206 & 1.516 & 0.051 & 39.812 & 32.541 \\
\textbf{GPT-4o} & 14.330 & 0.532 & 31.304 & 17.278 & 4.666 & 0.190 & 33.051 & 22.539 \\
\textbf{Average} & 23.595 & 0.825 & 32.950 & 17.186 & 2.893 & 0.109 & 37.501 & 27.591 \\
\midrule
\multicolumn{9}{c}{\textit{\textbf{Open-Source Models}}} \\
\textbf{Qwen2.5-72B-Instruct} & 0.549 & 0.325 & 1.979 & 0.897 & 11.984 & 0.584 & 18.554 & 27.642 \\
\textbf{Llama-3.1-70B-Instruct} & 20.440 & 0.758 & 31.604 & 17.813 & -11.888 & -0.410 & 39.047 & 36.416 \\
\textbf{DeepSeek-67B-Chat} & 28.307 & 0.891 & 37.219 & 17.944 & 9.480 & 0.309 & 41.369 & 26.261 \\
\textbf{Yi-1.5-34B-Chat} & 13.620 & 0.434 & 36.778 & 22.790 & 6.325 & 0.227 & 37.503 & 25.707 \\
\textbf{Qwen2.5-32B-Instruc} & 11.566 & 0.869 & 15.608 & 7.984 & 2.823 & 0.194 & 19.571 & 7.883 \\
\textbf{DeepSeek-V2-Lite (15.7B)} & 4.804 & 0.153 & 36.846 & 20.562 & -9.504 & -0.311 & 41.199 & 21.270 \\
\textbf{Yi-1.5-9B-Chat}  & 7.953 & 0.253 & 36.799 & 26.545 & -3.684 & -0.119 & 41.818 & 35.417 \\
\textbf{Llama-3.1-8B-Instruct} & 20.521 & 0.646 & 37.240 & 21.104 & 4.939 & 0.163 & 40.928 & 29.466 \\
\textbf{Qwen-2.5-Instruct-7B} & 19.477 & 0.612 & 37.289 & 20.796 & -1.339 & -0.075 & 24.124 & -16.053 \\
\textbf{Average} & 14.137 & 0.549 & 30.151 & 17.382 & 1.015 & 0.062 & 33.790 & 21.557 \\
\bottomrule
\end{tabular}
}
\end{threeparttable}
\end{table}

\begin{table}[t]
\centering
\caption{Performance of ETF trading with different LLMs as backbone models.}
\label{tab:etf-trading-performance}
\begin{threeparttable}
\scalebox{0.7}{  
\begin{tabular}{@{}lcccc@{}}  
\toprule
\textbf{ETF} & \textbf{CR\(\uparrow\)} & \textbf{SR\(\uparrow\)} & \textbf{AV\(\downarrow\)} & \textbf{MDD\(\downarrow\)} \\
\midrule
\textbf{Buy \& Hold} & 2.069 & 0.06 & 46.645 & 35.746 \\
\midrule
\multicolumn{4}{c}{\textit{\textbf{Financial Domain Models}}} \\
\textbf{Palmyra-Fin-70B} & 24.759 & 1.152 & 30.419 & 8.203 \\
\midrule
\multicolumn{4}{c}{\textit{\textbf{Proprietary Models}}} \\
\textbf{GPT-o1-preview} & 21.224 & 0.849 & 43.766 & 20.054 \\
\textbf{GPT-4} & 2.807 & 0.110 & 44.679 & 37.785 \\
\textbf{GPT-4o} & 12.292 & 0.377 & 46.150 & 32.678 \\
\textbf{Average} & 12.108 & 0.445 & 44.865 & 30.172 \\
\midrule
\multicolumn{4}{c}{\textit{\textbf{Open-Source Models}}} \\
\textbf{Qwen2.5-72B-Instruct} & 4.507 & 0.227 & 28.090 & 8.580 \\
\textbf{Llama-3.1-70B-Instruct} & 9.895 & 0.464 & 30.184 & 12.759 \\
\textbf{Yi-1.5-34B-Chat} & 4.996 & 0.322 & 21.986 & 12.858 \\
\textbf{Qwen2.5-32B-Instruct} & 19.617 & 0.955 & 29.070 & 7.496 \\
\textbf{DeepSeek-V2-Lite (15.7B)} & 1.389 & 0.063 & 31.371 & 31.831 \\
\textbf{Yi-1.5-9B-Chat} & -4.657 & -0.228 & 28.907 & 15.545 \\
\textbf{Llama-3.1-8B-Instruct} & 11.239 & 0.475 & 33.480 & 15.587 \\
\textbf{Qwen-2.5-Instruct-7B} & -0.384 & -0.020 & 27.596 & 14.059 \\
\textbf{Average} & 5.825 & 0.282 & 28.835 & 14.839 \\
\bottomrule
\end{tabular}
}
\end{threeparttable}
\label{etf_result}
\end{table}

\subsection{Discussion}
Combining all the experimental results, we find that the performance of different LLM varies significantly in stock, cryptocurrency, and ETF trading. This variation not only reflects the inherent complexity of financial markets, but also highlights the importance of model selection or fine-tuning. For instance, proprietary LLM generally exhibit be performance in stock trading due to their strong training on various financial datasets, while open-source models struggle to achieve these results, especially in more volatile environments such as cryptocurrency trading. In addition, the effectiveness of LLM-based agents depends heavily on their ability to adapt to market fluctuations. Agents that incorporate advanced memory systems and dynamic risk assessment capabilities are better able to cope with complex market situations, highlighting the value of the complex architectural features of LLM-based agent framework in financial decision-making tasks.

%% file: appendix.tex
\noindent {\large \textbf{Appendices}}
\section{Memory Ranking Mechanism of \textsc{FinMem}}
\label{sec:memory_ranking}
Upon receiving an investment inquiry, \textsc{FinMem} retrieves the top-$K$ critical memory events from each layer and channels them to the immediate reflection component of the working memory. These events are selected based on their information retrieval score, $\gamma_l^{E}$, where $l$ represents the layer (shallow, intermediate, or deep), as defined in Equation~\ref{eqn:eq_all}.

\begin{equation}
\gamma_l^{E} =   S_{\text{Recency}_{l}}^{E} + S_{\text{Relevancy}_{l}}^{E} + S_{\text{Importance}_{l}}^{E},
 \label{eqn:eq_all}
\end{equation}
where each memory event is only associated with one score and can only belong to a single layer.
\\

Let $E$ denote a given memory event. The scoring mechanism for $E$, adapted from Park et al. \cite{10.1145/3586183.3606763} but with modified recency and importance computations, is tailored to handle data with various timelines and to achieve layered processing that represents the diverse periodicities of the financial environment. This score encapsulates three metrics: recency (how recently the event occurred), relevancy (the event's pertinence to the current context), and importance (the event's significance). Individual metric scores exceeding 1.0 are scaled to the [0,1] range before being summed, ensuring a balanced contribution from each component and preventing any single metric from dominating the overall score. The resulting composite score provides a comprehensive evaluation of the memory event's significance within the multi-layered, periodically varying financial landscape.
\begin{equation}
  \begin{split}
& S_{\text{Recency}_{l}}^{E} = e^{-\frac{\delta^{E}}{Q_l}},  \quad \; \delta^{E} = t_{\text{P}} - t_{E}, \\
 \end{split}
 \label{eqn:eq1}
  \end{equation}

where  $\delta^{E}$ represents the time elapsed between a memory event's occurrence and the trading inquiry's arrival. The model utilizes three processing layers, each corresponding to a specific timeframe: shallow ($Q_{\text{shallow}} = 14$ days), intermediate ($Q_{\text{intermediate}} = 90$ days), and deep ($Q_{\text{deep}} = 365$ days). These intervals represent two weeks, a quarter, and a year respectively.
\\
 

When a trade inquiry $P$ arrives in processing layer $l$ via an LLM prompt, the agent calculates the recency score $S_{\text{Recency}_{l}}^{E}$ for a memory event $E$ using Equation \ref{eqn:eq1}. This score inversely correlates with the time elapsed between the inquiry and the event's memory timestamp, mapping to Ebbinghaus's forgetting curve \cite{murre2015replication}. The stability term $Q_l$ in Equation \ref{eqn:eq1} modulates memory decay rates across layers, with higher values in deeper layers indicating longer memory persistence. For instance, in the trading context, company annual reports (e.g., Form 10-Ks) are assigned higher stability values and categorized within deeper processing layers compared to daily financial news, reflecting their extended timeliness, relevance, and impact on financial decision-making. 

\begin{equation}
  \begin{split}
& S_{\text{Relevancy}_{l}}^{E} = \frac{\mathbf{m_{E}} \cdot \mathbf{m_{P}}}{\|\mathbf{m_{E}}\|_2 \times \|\mathbf{m_{P}}\|_2}
 \end{split}
 \label{eqn:eq2}
  \end{equation}

The relevancy score $S_{\text{relevancy}{l}}^{E}$ quantifies the semantic similarity between a memory event $E$ and the current query $P$ using cosine similarity of their respective embedding vectors, $\mathbf{m{E}}$ and $\mathbf{m_{P}}$, as shown in Equation \ref{eqn:eq2}. These embeddings are generated from the event's textual content and the LLM prompt query (which includes trading inquiries and the agent's character setting) using OpenAI's "text-embedding-ada-003" model.

The importance score $S_{\text{Importance}{l}}^{E}$ for a memory event $E$ in layer $l$ is calculated as the product of a value $v_l^{E}$ (derived from a uniform piecewise scoring function, Equation \ref{eqn:eq3}) and a degrading ratio $\theta_l$ (Equation \ref{eqn:eq4}), as shown in Equation \ref{eqn:eq5}. This approach, adapted from \cite{10.1145/3586183.3606763}, is tailored to our stratified long-term memory structure. The likelihood of higher $v_l^{E}$ values increases from shallow to deep layers, while $\theta_l$ measures the diminishing importance of an event over time using layer-specific exponential functions. The base $\alpha_l$ for each layer follows $\alpha{shallow} < \alpha_{intermediate} < \alpha_{deep}$ (set to 0.9, 0.967, and 0.988 respectively), ensuring $\theta_l$ decreases to a threshold of 5 after 30, 90, and 365 days for shallow, intermediate, and deep layers. This layered approach, implemented through three-piece-wise functions for both $S_{\text{Importance}{l}}^{E}$ and $S{\text{Recency}{l}}^{E}$, enables FinMem to process long-term memory in a stratified manner. Memory events are purged when $S{\text{Recency}{l}}^{E}$ falls below 0.05 or $S{\text{Importance}_{l}}^{E}$ is under 5 (pre-scaling), maintaining the relevance and efficiency of the memory store.

\begin{equation}
  \begin{split}
& v_{l}^{E} = \begin{cases} 
40 & \text{with probability } p_1\\
60 & \text{with probability } p_2\\
80 & \text{with probability } p_3
\end{cases}
 \end{split}
 \label{eqn:eq3}
  \end{equation}

\begin{equation}
  \begin{split}
    \theta_{l} = (\alpha_l)^{\delta^{E}}, \quad & l = \text{shallow}, \text{intermediate}, \text{deep},
  \end{split}
  \label{eqn:eq4}
\end{equation}

where $p_1 + p_2 + p_3 = 1$, but their values vary by shallow, intermediate, and deep processing. when shallow processing ${p_1, p_2, p_3} = \{0.8, 0.15, 0.05\}$, intermediate processing, ${p_1, p_2, p_3} = \{0.05, 0.8, 0.15\}$ and deep processing, ${p_1, p_2, p_3} = \{0.05, 0.15, 0.8\}$. 

\begin{equation}
S_{\text{Importance}_{l}}^{E} = v_{l}^{E} * \theta_{l},
\label{eqn:eq5}
\end{equation}

Furthermore, \textsc{FinMem} employs an access counter function to dynamically manage memory events across layers, ensuring that crucial events influencing trading decisions are elevated to deeper layers for extended retention and recurring access. This process, monitored by the LLM validation tool Guardrails AI, tracks critical memory IDs across layers. Events deemed pivotal for investment success receive a 5-point boost to their importance score ($S_{\text{Importance}{l}}^{E}$). Upon meeting upgrade criteria for a deeper layer, an event's recency score ($S{\text{Recency}_{l}}^{E}$) is reset to 1.0, underscoring its significance and preventing rapid decay. Conversely, less relevant events gradually fade. This mechanism allows \textsc{FinMem} to efficiently identify, prioritize, and retain key events based on their nature and retrieval frequency, while gradually phasing out less impactful information, thereby maintaining a dynamic and relevant memory structure for financial decision-making.

\section{Details on Evaluation Metrics}
\label{sec:metric}
Below is a brief overview of these metrics:\par
\noindent\textbf{Cumulative Return (CR) \%} measures the total value change of an investment over time by summing daily logarithmic returns, shown in Equation~\ref{eq:cum_return}. Higher values indicate better strategy effectiveness.
\begin{align}
   \label{eq:cum_return}
   \textbf{CR} &= \sum_{t=1}^{n} r_i = \sum_{t=1}^{n} \left[ \ln\left(\frac{p_{t+1}}{p_t}\right) \cdot \text{action}_t \right]
\end{align}

, where $r_i$ is the logarithmic return from day $t$ to $t+1$, $p_t$ and $p_{t+1}$ are the closing prices on days $t$ and $t+1$, respectively, and $\text{action}_t$ is the model's trading decision for day $t$.

\noindent\textbf{Sharpe Ratio (SR)} assesses risk-adjusted returns by dividing the average excess return ($R_p$) over the risk-free rate ($R_f$) by its volatility ($\sigma_p$), detailed in Equation~\ref{eq:sharpe}. Higher ratios signify better performance.
  
\begin{equation}
    \textbf{SR} = \frac{R_p - R_f}{\sigma_p}
    \label{eq:sharpe}
\end{equation}  

\noindent\textbf{Annualized Volatility (AV) \% and Daily Volatility (DV) \%} quantify return fluctuations; AV is derived by scaling DV (\textit{standard deviation of daily logarithmic returns}) by the square root of the annual trading days (252), as in Equation~\ref{eq:annuaVol}. This metric highlights potential return deviations across the year.
\begin{align}
   \label{eq:annuaVol}
    \textbf{AV} &= \textbf{DV} \times \sqrt{252} 
\end{align} 

\noindent\textbf{Max Drawdown (MDD) \%} calculates the largest portfolio value drop from peak to trough, as given in Equation~\ref{eq:maxdrawdown}. Lower values indicate lesser risk and higher strategy robustness. 
    \begin{align}
    \label{eq:maxdrawdown}
    \textbf{MDD} = \text{max}(\frac{P_{\text{peak}} - P_{\text{trough}}}{P_{\text{peak}}})
    \end{align}

Note that CR and the SR are often considered more essential than AV and MDD in evaluating asset trading performance due to their focus on long-term gains and risk-adjusted returns by their definition. Here, we regard these two metrics as primary metrics when evaluating the experiment outcomes.

\section{An example of mixed and lagged market signals: Partial investment insights of TSLA on 2021-03-05} 
\label{sec:example_tsla}
Here are some insights from the agent's memory module for TSLA as of 2021-03-05. A few memory records are omitted; these are either neutral or positive. Despite this, the stock price trend for TSLA is sharply downward, conflicting with the overall positive financial sentiments and market signals. Utilizing proprietary models such as GPT4 and GPT-o1 as backbones, the financial decision-making agent can leverage other investment insights like historical momentum and self-reflection to consistently support a 'Sell' decision. In contrast, the large-sized open-source models like Qwen2.5-72B and DeepSeek-67B-Chat exhibit instability in producing consistent actions across repeated experimental trials.

\noindent\textbf{Short-term Memory}
\begin{itemize}
    \item[1] \textbf{Sentiment: Negative}: The key insights from the news regarding Tesla Inc (NASDAQ: TSLA) losing market share to Ford Motor Company's (NYSE: F) Mustang Mach-E in the United States are as follows:1. **Market Competition**: Tesla is facing increased competition in the electric vehicle (EV) market, particularly from established automakers like Ford. The Mustang Mach-E's success indicates that other companies are effectively entering the EV space, which could impact Tesla's dominance.2. **Market Share Impact**: The loss of market share to Ford suggests that Tesla's growth trajectory in the U.S. may be challenged. Investors should consider the potential implications for Tesla's revenue and profitability if this trend continues...
    \item[2] \textbf{Sentiment: Positive}: The news about a Banksy artwork being burned and sold as a non-fungible token (NFT) highlights a few key insights relevant to investment decisions, particularly concerning innovative and disruptive technologies: 1. **Emergence of NFTs**: The transformation of physical art into digital assets through NFTs signifies a growing trend in the digital economy. This trend could influence sectors beyond art, including technology and finance, as more industries explore blockchain applications ...
    \item[3] \textbf{Sentiment: Neutral}: To provide a summary of key insights relevant to making investment decisions about Tesla (TSLA) from the pre-open movers news, I would focus on the following aspects: 1. **Stock Performance**: Look for any significant pre-market price movements for TSLA. If the stock is showing a notable increase or decrease, it could indicate investor sentiment or reaction to recent news.2. **News Catalysts**: Identify any specific news items or announcements related to Tesla that might be influencing its stock price. This could include earnings reports, product launches, regulatory news, or changes in leadership...
    \item[4] \textbf{Sentiment: positive}: The key insights from the news about Tesla's vehicle registrations in Germany are as follows: 1. **Significant Growth in Registrations**: Tesla experienced a 78\% year-over-year increase in vehicle registrations in Germany as of January. This substantial growth indicates a strong demand for Tesla vehicles in one of Europe's largest automotive markets.2. **Market Penetration**: The surge in registrations suggests that Tesla is successfully penetrating the German market, which is traditionally dominated by local automakers. This could imply a growing acceptance and preference for electric vehicles (EVs) in Germany, benefiting Tesla as a leading EV manufacturer...
\end{itemize}
...

\noindent\textbf{Mid-term Memory}
\begin{itemize}
    \item[1] \textbf{Sentiment: Negative}: The news about U.S. Senator Elizabeth Warren's proposal for a Democratic wealth tax could have several implications for investment decisions regarding Tesla (TSLA):1. **Impact on Wealthy Investors**: The proposed wealth tax targets the wealthiest Americans, which could include major shareholders and investors in Tesla. If implemented, this tax could lead to changes in investment strategies among these individuals, potentially affecting Tesla's stock price due to shifts in large-scale holdings....5. **Long-term Implications**: While the immediate impact might be uncertain, the long-term implications of a wealth tax could lead to a more equitable distribution of wealth, potentially increasing consumer spending power. This could benefit Tesla if it results in a larger market for its products. Investors in Tesla should closely monitor the progress of this proposal and consider its potential effects on market dynamics and investor behavior. 
    \item[2] \textbf{Sentiment: positive}: The key insights from the news information regarding consumer preferences in China for electric cars, particularly focusing on price and driving range, can be relevant for making investment decisions about Tesla (TSLA) in the following ways: 1. **Market Demand Alignment**: Tesla's ability to align its product offerings with the primary concerns of Chinese consumers 2014price and driving range\u2014could significantly impact its market share and sales growth in China. Investors should monitor how Tesla's models compare to competitors in these aspects... Overall, Tesla's success in addressing the key consumer preferences in China 2014price and driving range 2014will be critical for its growth prospects in the region, making these factors important considerations for investors. 
\end{itemize}
...

\noindent\textbf{Long-term Memory}
\begin{itemize}
    \item[1] \textbf{Sentiment: Negative}: The key insights from the news about Tesla Inc. (NASDAQ: TSLA) facing competition in China are as follows: 1. **Increased Competition**: Tesla's Model 3 is experiencing significant competition from a budget electric vehicle produced through a joint venture involving General Motors (NYSE: GM). This indicates a more competitive landscape in the Chinese electric vehicle (EV) market, which could impact Tesla's market share and sales growth in the region... For investors, these insights suggest a need to closely watch Tesla's strategic responses to competition in China, its pricing strategies, and any potential impacts on its financial performance. Additionally, understanding the broader competitive landscape and Tesla's ability to sustain its growth trajectory will be crucial for making informed investment decisions. 
    \item[2] \textbf{Sentiment: positive}: The news about Bill Gates' concerns regarding Bitcoin's impact on climate change highlights a broader issue of environmental sustainability in the tech and financial sectors. Here are the key insights relevant to making investment decisions about Tesla (TSLA): 1. **Environmental Impact Awareness**: Bill Gates' concerns underscore the growing awareness and scrutiny of the environmental impact of technology and financial products. This is relevant for Tesla, as the company positions itself as a leader in sustainable energy and electric vehicles (EVs)... Overall, the emphasis on environmental impact and sustainability in the tech sector could reinforce Tesla's strategic advantages and appeal to investors prioritizing green investments. 
\end{itemize}
...

\section{Case Study on Single Stock Trading: Forecast for TSLA on 2022-10-25 to Predict Trading Decision on 2022-10-26} 
\label{sec:user_case}

\begin{figure*}[ht!]
\centering
\begin{tcbraster}[raster columns=1] 
    \begin{tcolorbox}[colback=white, colframe=lightgrey, title=Initialize Profile, boxsep=2pt, top=2pt, bottom=2pt]
    \small
    \raggedright
    \textbf{\textcolor{lightgrey}{1. Operations:}} \\
    - Provide a performance overview of the trading stock based on available data.\\
    - Set up the risk inclination as the key character of the trading agent.\\
    \textbf{\textcolor{lightgrey}{2. Range:}} Financial information such as the financial sectors, historical performance, and previous stock trends of the trading stock.\\
    \textbf{\textcolor{lightgrey}{3. Prompts:}} You are an experienced trading manager and investment firm. Your task is to make informed decisions on the given stock based on the provided information.\\
    \textcolor{red}{Under Self-Adaptive Risk Character Setting:} When historical momentum is positive, you are a risk-seeking investor. But when historical momentum is negative, you are a risk-averse investor.\\

    \textbf{\textcolor{lightgrey}{4. General background setting:}} \\
    You have accumulated a lot of information about the following sectors, so you are especially good at trading them: 1)Electric Vehicles (Automotive Sector). 2) Energy Generation and Storage...From year 2021 to 2022 September, Tesla's continued growth and solid financial performance over the defined period ...\\
    \end{tcolorbox}
    \begin{tcolorbox}[colback=white, colframe=lightblue, title=Summarize, boxsep=2pt, top=2pt, bottom=2pt]
    \small
    \raggedright
    \textbf{\textcolor{lightblue}{1. Operations:}} \\
    - Summarize different types of input information. \\
    - Distribute them to corresponding layers of the long-term memory database.\\
    \textbf{\textcolor{lightblue}{2. Range:}} Daily market news, Long Documents such as company 10-K and 10-Q reports \\
    \textbf{\textcolor{lightblue}{3. Prompts:}}\\
    - (1). Summarize the contents: Summarize the following documents into {1000} words.\\
    - (2). Comprehend the investment sentiment of news insights: The positive, neutral and negative scores are for understanding the investment sentiments, opinions, or emotions. For example, positive news about a company can lift investor sentiment, encouraging more buying activity, which in turn can push stock prices higher...\\
    \textbf{\textcolor{lightblue}{4. Outputs:}} \\
    (1). \textbf{To Shallow Memory Layer:} \\
    - [News \textcolor{darkblue}{(ID: 261)}] Here's How Much You Would Have Made Owning Tesla Stock In The Last 10 Years Tesla (NASDAQ:TSLA) has outperformed the market over the past 10 years by 50.69\% on an annualized basis producing an average annual return of 60.76\%. Currently, Tesla has a market capitalization of \$683.54 billion.... The sentiment is \textcolor{red}{\{positive\}}. \\
    - [News \textcolor{darkblue}{(ID: 278)}] Tesla Q3 Earnings Are Imminent. Can Nio Foreshadow What's To Come? What To Know Before The Print Tesla Inc (NASDAQ: TSLA) shares were trading down slightly Wednesday afternoon ahead of the automaker\'s third-quarter report, but the stock is up 6\% over the last five sessions... The sentiment is \textcolor{red}{\{positive\}}.\\
    - ... \\
    (2). \textbf{To Intermediate Memory Layer:} \\ 
    - [Form 10-Q \textcolor{darkblue}{(ID: 222)}] Tesla Q3 2022 revenues were \$21.5 billion, up 56\% year-over-year. Automotive sales revenue grew 56\% to \$17.8 billion driven by higher Model 3/Y and Model S/X deliveries. Gross automotive margin declined to 27.9\% due to cost inflation and factory ramps. Net income was \$3.3 billion, up 102\% year-over-year. Positive free cash flow was \$6.1 billion...  \\
    - [News \textcolor{darkblue}{(ID: 275)}] Tesla Q3 Earnings Highlights: Record Revenue, Operating Margin And Free Cash Flow, Tesla Semi Deliveries Coming In December Electric vehicle leader Tesla Inc (NASDAQ: TSLA) reported third-quarter financial results after market close Wednesday...The sentiment is \textcolor{red}{\{neutral\}}.  \\
    - [News \textcolor{darkblue}{(ID: 274)}] Tesla Preps For 2023 Cybertruck Launch, Will Make Battery Packs In California The Cybertruck is one of Tesla Inc. (NASDAQ: TSLA) most hotly anticipated, but also most delayed, products.
    - ...The sentiment is \textcolor{red}{\{negative\}}.\\
    (3). \textbf{To Deep Memory Layer:} \\ 
    - [News \textcolor{darkblue}{(ID: 161)}] Tesla Whale Trades Spotted A whale with a lot of money to spend has taken a noticeably bearish stance on Tesla. Looking at the options history for Tesla (NASDAQ:TSLA) we detected 477 strange trades. The sentiment is \textcolor{red}{\{positive\}}.\\
    - [Self-reflection \textcolor{darkblue}{(ID: 226)}] Given the short-term positive news score in the market for TSLA and a positive cumulative return, there is a high probability of continued growth in the short term. However, investor should be aware of potential threats in the mid-term market with competitors like General Motors, and Nio...
    \end{tcolorbox}
    \begin{tcolorbox}[colback=white, colframe=lightpurple, title=Observe, boxsep=2pt, top=2pt, bottom=2pt]
    \small
    \raggedright
    \textbf{\textcolor{lightpurple}{1. Operations:}} Access and interpret market indicators such as current stock prices and historical momentum data. \\
    \textbf{\textcolor{lightpurple}{2. Range:}} Stock's daily adjusted closing price, historical momentum in the past $k$ days ($k=3$ in this case), etc. \\
    \textbf{\textcolor{lightpurple}{3. Prompts:}}\\
    - The information below provides a summary of stock price fluctuations over the previous few days, which is the "momentum" of a stock. It reflects the trend of a stock. Momentum is based on the idea that securities that have performed well in the past will continue to perform well, and conversely, securities that have performed poorly will continue to perform poorly.\\
    \textbf{\textcolor{lightpurple}{4. Outputs:}}\\
    - (1). The daily adjusted closing price of TSLA on \textcolor{red}{\{2022-10-25\}} is \textcolor{red}{\{\$222.42\}}.\\
    - (2). \textbf{Train:} On \textcolor{red}{\{2022-10-25\}}, the momentum of TSLA, indicated by the price difference between the current and the next trading \\
    \quad \quad \; day, is \textcolor{red}{\{\$2.22\}}.\\
    \quad \quad \; \textbf{Test:} On \textcolor{red}{\{2022-10-25\}}, the historical momentum of TSLA, as measured by its cumulative logarithmic returns in the past \\
    \quad \quad \; \textcolor{red}{\{3\}} days, was \textcolor{red}{\{7.05\%\}}. \\
    \end{tcolorbox}
\end{tcbraster}
\caption{\small First section of \textsc{FinMem}'s workflow for perceiving and processing multi-sourced information from market environment.}
\label{fig_casestudy:modular_design_p1}
\end{figure*}

\begin{figure*}[ht!]
\centering
\begin{tcbraster}[raster columns=1, raster equal height] 
    \begin{tcolorbox}[colback=white, colframe=lightorange, title=Reflect, boxsep=2pt, top=2pt, bottom=2pt]
    \small
    \raggedright
    \textbf{\textcolor{lightorange}{1. Operations:}}\\
    \textbf{Train:} - Infer the reasoning from the retrieved memories insights for the recommended trading actions the increase or decrease of the trading stock price.\\
    \textbf{Test:} - Generate recommended trading actions. \quad \quad - Infer the reasoning for making such trading action. \\
    \textbf{Train \& Test:} - Track the memory insight IDs that contributed to the current trading decision.\\
    \textbf{\textcolor{lightorange}{2. Range:}} Interpretation of observed market indicators and top $k$ insights from each layer of the long-term memory database. \\
    \textbf{\textcolor{lightorange}{3. Prompts:}}\\
    \textbf{\textit{Train:}} Given the following information, can you explain why the financial market fluctuation from the current day to the next day behaves like this? Summarize the reason for the decision. You should provide summary information and the IDs of the information to support your summary.\\
    \textbf{\textit{Test:}} Given the information, can you make an investment decision? Just summarize the reason for the decision. Consider only the available short-term information, the mid-term information, the long-term information, and the reflection-term information.
    Consider the momentum of the historical stock price.
    Also, consider how much share of the stock the investor holds now. You should provide exactly one of the following investment decisions: buy or sell.
    You could go with the 'hold' option if necessary.
    You also need to provide the IDs of the information to support your decision.\\
    \textbf{\textcolor{lightorange}{4. Outputs:}}\\
    \textbf{Train:}\\
    - \textcolor{red}{Recommended Action Reason:}  The recommended trading action is 'Buy'. This can be attributed to positive market sentiment driven by a range of favorable factors. Notably, Tesla's strong battery supply chain, effective market expansion plans in Germany, and its ability to maintain market strength despite potential competitive threats positively influenced investor confidence, contributing to the stock price rise.\\
    - \textcolor{red}{Reflected Contributed Insights IDs:} \textcolor{red}{\{261, 278, 275, 274, 161,...\}}\\
    \textbf{Test:}\\
    - \textcolor{red}{Trading Action: \{Buy\}}\\   
    - \textcolor{red}{Recommended Action Reason:} Considering the short-term positive revenue growth, higher Model 3/Y and Model S/X deliveries, positive free cash flow, and increased cash and cash equivalents, Tesla appears to have strong financial health. The momentum being positive, it also encourages a risk-seeking strategy. The significant positive signs from news such as the imminent Q3 earnings add to the favorable view of the investment. Despite the mid-term neutral sentiment, long-term information shows only marginal drops and the stock is still looking strong overall. These, along with positive reflection-term information, form the reasoning behind the 'buy' decision...\\ 
    - \textcolor{red}{Reflected Contributed Insights IDs: \{261, 278, 275, 274, 161,...\}}\\
    \end{tcolorbox}
\end{tcbraster}
\caption{\small Second section of \textsc{FinMem}'s workflow for generating trading action, reasoning and reflection.}
\label{fig_casestudy:modular_design_p2}
\end{figure*}

%% file: main.bbl
\begin{thebibliography}{43}
\providecommand{\natexlab}[1]{#1}

\bibitem[{Achiam et~al.(2023)Achiam, Adler, Agarwal, Ahmad, Akkaya, Aleman, Almeida, Altenschmidt, Altman, Anadkat et~al.}]{achiam2023gpt}
Josh Achiam, Steven Adler, Sandhini Agarwal, Lama Ahmad, Ilge Akkaya, Florencia~Leoni Aleman, Diogo Almeida, Janko Altenschmidt, Sam Altman, Shyamal Anadkat, et~al. 2023.
\newblock Gpt-4 technical report.
\newblock \emph{arXiv preprint arXiv:2303.08774}.

\bibitem[{Ang and Chen(2003)}]{ang2003downside}
Andrew Ang and Joseph Chen. 2003.
\newblock Downside risk.
\newblock \emph{Journal of Portfolio Management}, 29(4):103--112.

\bibitem[{Araci(2019)}]{araci2019finbert}
Dogu Araci. 2019.
\newblock Finbert: Financial sentiment analysis with pre-trained language models.
\newblock \emph{arXiv preprint arXiv:1908.10063}.

\bibitem[{Bertsekas and Shreve(1996)}]{bertsekas1996stochastic}
Dimitri Bertsekas and Steven~E Shreve. 1996.
\newblock \emph{Stochastic optimal control: the discrete-time case}, volume~5.
\newblock Athena Scientific.

\bibitem[{Bhatnagar et~al.(2023)Bhatnagar, Taneja, and Rupeika-Apoga}]{bhatnagar2023demystifying}
Mukul Bhatnagar, Sanjay Taneja, and Ramona Rupeika-Apoga. 2023.
\newblock Demystifying the effect of the news (shocks) on crypto market volatility.
\newblock \emph{Journal of Risk and Financial Management}, 16(2):136.

\bibitem[{Cochrane(1988)}]{cochrane1988volatility}
John~H. Cochrane. 1988.
\newblock Volatility tests and efficient markets: A review essay.
\newblock \emph{Journal of Monetary Economics}, 22(3):463--485.

\bibitem[{Eigner and H{\"a}ndler(2024)}]{eigner2024determinants}
Eva Eigner and Thorsten H{\"a}ndler. 2024.
\newblock Determinants of llm-assisted decision-making.
\newblock \emph{arXiv preprint arXiv:2402.17385}.

\bibitem[{Griffiths et~al.(2023)Griffiths, Zhu, Grant, and McCoy}]{griffiths2023bayes}
Thomas~L Griffiths, Jian-Qiao Zhu, Erin Grant, and R~Thomas McCoy. 2023.
\newblock Bayes in the age of intelligent machines.
\newblock \emph{arXiv preprint arXiv:2311.10206}.

\bibitem[{Guo et~al.(2024)Guo, Chen, Wang, Chang, Pei, Chawla, Wiest, and Zhang}]{guo2024large}
Taicheng Guo, Xiuying Chen, Yaqi Wang, Ruidi Chang, Shichao Pei, Nitesh~V Chawla, Olaf Wiest, and Xiangliang Zhang. 2024.
\newblock Large language model based multi-agents: A survey of progress and challenges.
\newblock \emph{arXiv preprint arXiv:2402.01680}.

\bibitem[{Huang et~al.(2023)Huang, Wang, and Yang}]{huang2023finbert}
Allen~H Huang, Hui Wang, and Yi~Yang. 2023.
\newblock Finbert: A large language model for extracting information from financial text.
\newblock \emph{Contemporary Accounting Research}, 40(2):806--841.

\bibitem[{Hull(2007)}]{hull2007risk}
John Hull. 2007.
\newblock \emph{Risk Management and Financial Institutions}.
\newblock John Wiley \& Sons.

\bibitem[{Kabbani and Duman(2022)}]{kabbani2022deep}
Taylan Kabbani and Ekrem Duman. 2022.
\newblock Deep reinforcement learning approach for trading automation in the stock market.
\newblock \emph{IEEE Access}, 10:93564--93574.

\bibitem[{Li et~al.(2024)Li, Luo, Wang, Chen, Liu, and He}]{li2024cryptotrade}
Yuan Li, Bingqiao Luo, Qian Wang, Nuo Chen, Xu~Liu, and Bingsheng He. 2024.
\newblock Cryptotrade: A reflective llm-based agent to guide zero-shot cryptocurrency trading.
\newblock In \emph{Proceedings of the 2024 Conference on Empirical Methods in Natural Language Processing}, pages 1094--1106.

\bibitem[{Liu et~al.(2023)Liu, Wang, and Zha}]{liu2023fingpt}
Xiao-Yang Liu, Guoxuan Wang, and Daochen Zha. 2023.
\newblock Fingpt: Democratizing internet-scale data for financial large language models.
\newblock \emph{arXiv preprint arXiv:2307.10485}.

\bibitem[{Liu et~al.(2020)Liu, Liu, Zhao, Pan, and Liu}]{liu2020adaptive}
Yang Liu, Qi~Liu, Hongke Zhao, Zhen Pan, and Chuanren Liu. 2020.
\newblock Adaptive quantitative trading: An imitative deep reinforcement learning approach.
\newblock In \emph{Proceedings of the AAAI conference on artificial intelligence}, volume~34, pages 2128--2135.

\bibitem[{Liu et~al.(2021)Liu, Huang, Huang, Li, and Zhao}]{liu2021finbert}
Zhuang Liu, Degen Huang, Kaiyu Huang, Zhuang Li, and Jun Zhao. 2021.
\newblock Finbert: A pre-trained financial language representation model for financial text mining.
\newblock In \emph{Proceedings of the twenty-ninth international conference on international joint conferences on artificial intelligence}, pages 4513--4519.

\bibitem[{Llama~team(2024)}]{llama3.1}
Meta Llama~team. 2024.
\newblock \href {https://ai.meta.com/research/publications/the-llama-3-herd-of-models/} {The llama 3 herd of models}.

\bibitem[{Madhavan(2016)}]{madhavan2016exchange}
Ananth~N Madhavan. 2016.
\newblock \emph{Exchange-traded funds and the new dynamics of investing}.
\newblock Oxford University Press.

\bibitem[{Murre and Dros(2015)}]{murre2015replication}
Jaap~MJ Murre and Joeri Dros. 2015.
\newblock Replication and analysis of ebbinghaus’ forgetting curve.
\newblock \emph{PloS one}, 10(7):e0120644.

\bibitem[{OpenAI(2022)}]{ChatGPT}
OpenAI. 2022.
\newblock \href {https://openai.com/blog/chatgpt} {Introducing chatgpt}.

\bibitem[{Park et~al.(2023)Park, O'Brien, Cai, Morris, Liang, and Bernstein}]{10.1145/3586183.3606763}
Joon~Sung Park, Joseph O'Brien, Carrie~Jun Cai, Meredith~Ringel Morris, Percy Liang, and Michael~S. Bernstein. 2023.
\newblock \href {https://doi.org/10.1145/3586183.3606763} {Generative agents: Interactive simulacra of human behavior}.
\newblock In \emph{Proceedings of the 36th Annual ACM Symposium on User Interface Software and Technology}, UIST '23, New York, NY, USA. Association for Computing Machinery.

\bibitem[{Qwen~team(2024)}]{Qwen2.5}
Alibaba~Group Qwen~team. 2024.
\newblock \href {https://qwen.readthedocs.io/en/latest/} {Qwen2.5}.

\bibitem[{Saqur et~al.(2024)Saqur, Kato, Vinden, and Rudzicz}]{saqur2024nifty}
Raeid Saqur, Ken Kato, Nicholas Vinden, and Frank Rudzicz. 2024.
\newblock Nifty financial news headlines dataset.
\newblock \emph{arXiv preprint arXiv:2405.09747}.

\bibitem[{Shah et~al.(2022)Shah, Chawla, Eidnani, Shah, Du, Chava, Raman, Smiley, Chen, and Yang}]{shah2022flue}
Raj~Sanjay Shah, Kunal Chawla, Dheeraj Eidnani, Agam Shah, Wendi Du, Sudheer Chava, Natraj Raman, Charese Smiley, Jiaao Chen, and Diyi Yang. 2022.
\newblock When flue meets flang: Benchmarks and large pre-trained language model for financial domain.
\newblock \emph{arXiv preprint arXiv:2211.00083}.

\bibitem[{Sharpe(1994)}]{sharpe1994sharpe}
William~F. Sharpe. 1994.
\newblock The sharpe ratio.
\newblock \emph{The Journal of Portfolio Management}, 21(1):49--58.

\bibitem[{team(2024)}]{Palmyra-Fin-70B-32k}
Writer~Engineering team. 2024.
\newblock {Palmyra-Fin-70B-32k: a powerful LLM designed for Finance}.
\newblock \url{https://dev.writer.com}.

\bibitem[{Vanhoucke(2023)}]{kagglecryptonews}
Olivier Vanhoucke. 2023.
\newblock \href {https://www.kaggle.com/datasets/oliviervha/crypto-news} {Crypto news dataset}.
\newblock Accessed: 2024-08-20.

\bibitem[{Wang et~al.(2024)Wang, Ma, Feng, Zhang, Yang, Zhang, Chen, Tang, Chen, Lin et~al.}]{wang2024survey}
Lei Wang, Chen Ma, Xueyang Feng, Zeyu Zhang, Hao Yang, Jingsen Zhang, Zhiyuan Chen, Jiakai Tang, Xu~Chen, Yankai Lin, et~al. 2024.
\newblock A survey on large language model based autonomous agents.
\newblock \emph{Frontiers of Computer Science}, 18(6):186345.

\bibitem[{Wu et~al.(2023)Wu, Irsoy, Lu, Dabravolski, Dredze, Gehrmann, Kambadur, Rosenberg, and Mann}]{wu2023bloomberggpt}
Shijie Wu, Ozan Irsoy, Steven Lu, Vadim Dabravolski, Mark Dredze, Sebastian Gehrmann, Prabhanjan Kambadur, David Rosenberg, and Gideon Mann. 2023.
\newblock Bloomberggpt: A large language model for finance.
\newblock \emph{arXiv preprint arXiv:2303.17564}.

\bibitem[{Xie et~al.(2024)Xie, Han, Chen, Xiang, Zhang, He, Xiao, Li, Dai, Feng et~al.}]{xie2024finben}
Qianqian Xie, Weiguang Han, Zhengyu Chen, Ruoyu Xiang, Xiao Zhang, Yueru He, Mengxi Xiao, Dong Li, Yongfu Dai, Duanyu Feng, et~al. 2024.
\newblock The finben: An holistic financial benchmark for large language models.
\newblock \emph{arXiv preprint arXiv:2402.12659}.

\bibitem[{Xie et~al.(2023)Xie, Han, Zhang, Lai, Peng, Lopez-Lira, and Huang}]{xie2023pixiu}
Qianqian Xie, Weiguang Han, Xiao Zhang, Yanzhao Lai, Min Peng, Alejandro Lopez-Lira, and Jimin Huang. 2023.
\newblock Pixiu: A large language model, instruction data and evaluation benchmark for finance.
\newblock \emph{arXiv preprint arXiv:2306.05443}.

\bibitem[{Xin et~al.(2024)Xin, Ren, Song, Shao, Zhao, Wang, Liu, Zhang, Lu, Du, Gao, Zhu, Yang, Gou, Wu, Luo, and Ruan}]{xin2024deepseek}
Huajian Xin, Z.~Z. Ren, Junxiao Song, Zhihong Shao, Wanjia Zhao, Haocheng Wang, Bo~Liu, Liyue Zhang, Xuan Lu, Qiushi Du, Wenjun Gao, Qihao Zhu, Dejian Yang, Zhibin Gou, Z.~F. Wu, Fuli Luo, and Chong Ruan. 2024.
\newblock \href {https://arxiv.org/abs/2408.08152} {Deepseek-prover-v1.5: Harnessing proof assistant feedback for reinforcement learning and monte-carlo tree search}.

\bibitem[{Yang et~al.(2024)Yang, Zhang, Wang, Guo, Zhang, Lin, Wang, Zhou, Guan, Zhang et~al.}]{yang2024finrobot}
Hongyang Yang, Boyu Zhang, Neng Wang, Cheng Guo, Xiaoli Zhang, Likun Lin, Junlin Wang, Tianyu Zhou, Mao Guan, Runjia Zhang, et~al. 2024.
\newblock Finrobot: An open-source ai agent platform for financial applications using large language models.
\newblock \emph{arXiv preprint arXiv:2405.14767}.

\bibitem[{Yang et~al.(2023)Yang, Tang, and Tam}]{yang2023investlm}
Yi~Yang, Yixuan Tang, and Kar~Yan Tam. 2023.
\newblock Investlm: A large language model for investment using financial domain instruction tuning.
\newblock \emph{arXiv preprint arXiv:2309.13064}.

\bibitem[{Yang et~al.(2020)Yang, Uy, and Huang}]{yang2020finbert}
Yi~Yang, Mark Christopher~Siy Uy, and Allen Huang. 2020.
\newblock Finbert: A pretrained language model for financial communications.
\newblock \emph{arXiv preprint arXiv:2006.08097}.

\bibitem[{Yi et~al.(2022)Yi, Xiao, Ijeoma, Cheran, Baweja, and Nelson}]{yi2022stock2vec}
Ziruo Yi, Ting Xiao, Kaz-Onyeakazi Ijeoma, Ratnam Cheran, Yuvraj Baweja, and Phillip Nelson. 2022.
\newblock Stock2vec: An embedding to improve predictive models for companies.
\newblock \emph{arXiv preprint arXiv:2201.11290}.

\bibitem[{Young et~al.(2024)Young, Chen, Li, Huang, Zhang, Zhang, Li, Zhu, Chen, Chang et~al.}]{young2024yi}
Alex Young, Bei Chen, Chao Li, Chengen Huang, Ge~Zhang, Guanwei Zhang, Heng Li, Jiangcheng Zhu, Jianqun Chen, Jing Chang, et~al. 2024.
\newblock Yi: Open foundation models by 01. ai.
\newblock \emph{arXiv preprint arXiv:2403.04652}.

\bibitem[{Yu et~al.(2024{\natexlab{a}})Yu, Li, Chen, Jiang, Li, Zhang, Liu, Suchow, and Khashanah}]{yu2024finmem}
Yangyang Yu, Haohang Li, Zhi Chen, Yuechen Jiang, Yang Li, Denghui Zhang, Rong Liu, Jordan~W Suchow, and Khaldoun Khashanah. 2024{\natexlab{a}}.
\newblock Finmem: A performance-enhanced llm trading agent with layered memory and character design.
\newblock In \emph{Proceedings of the AAAI Symposium Series}, volume~3, pages 595--597.

\bibitem[{Yu et~al.(2024{\natexlab{b}})Yu, Yao, Li, Deng, Cao, Chen, Suchow, Liu, Cui, Xu et~al.}]{yu2024fincon}
Yangyang Yu, Zhiyuan Yao, Haohang Li, Zhiyang Deng, Yupeng Cao, Zhi Chen, Jordan~W Suchow, Rong Liu, Zhenyu Cui, Zhaozhuo Xu, et~al. 2024{\natexlab{b}}.
\newblock Fincon: A synthesized llm multi-agent system with conceptual verbal reinforcement for enhanced financial decision making.
\newblock \emph{arXiv preprint arXiv:2407.06567}.

\bibitem[{Zhang et~al.(2024{\natexlab{a}})Zhang, Zhao, Xia, Sun, Sun, Qin, Li, Zhao, Zhao, Cai et~al.}]{zhang2024finagent}
Wentao Zhang, Lingxuan Zhao, Haochong Xia, Shuo Sun, Jiaze Sun, Molei Qin, Xinyi Li, Yuqing Zhao, Yilei Zhao, Xinyu Cai, et~al. 2024{\natexlab{a}}.
\newblock Finagent: A multimodal foundation agent for financial trading: Tool-augmented, diversified, and generalist.
\newblock \emph{arXiv preprint arXiv:2402.18485}.

\bibitem[{Zhang et~al.(2024{\natexlab{b}})Zhang, Mao, Ge, Wang, de~Wynter, Xia, Wu, Song, Lan, and Wei}]{zhang2024llm}
Yadong Zhang, Shaoguang Mao, Tao Ge, Xun Wang, Adrian de~Wynter, Yan Xia, Wenshan Wu, Ting Song, Man Lan, and Furu Wei. 2024{\natexlab{b}}.
\newblock Llm as a mastermind: A survey of strategic reasoning with large language models.
\newblock \emph{arXiv preprint arXiv:2404.01230}.

\bibitem[{Zhang and Zhang(2023)}]{zhang2023finbert}
Yuzhe Zhang and Hong Zhang. 2023.
\newblock Finbert--mrc: financial named entity recognition using bert under the machine reading comprehension paradigm.
\newblock \emph{Neural Processing Letters}, 55(6):7393--7413.

\bibitem[{Zhou et~al.(2021)Zhou, Ma, and Liu}]{zhou-etal-2021-trade}
Zhihan Zhou, Liqian Ma, and Han Liu. 2021.
\newblock \href {https://doi.org/10.18653/v1/2021.findings-acl.186} {Trade the event: Corporate events detection for news-based event-driven trading}.
\newblock In \emph{Findings of the Association for Computational Linguistics: ACL-IJCNLP 2021}, pages 2114--2124, Online. Association for Computational Linguistics.

\end{thebibliography}
